\documentclass[%
 reprint,
superscriptaddress,
 amsmath,amssymb,
aps,
prb,
floatfix,
]{revtex4-2}
\usepackage{silence}
\usepackage{chemfig}
\usepackage{ulem}
\usepackage{amsmath}
\usepackage{amssymb}
\usepackage{color}
\usepackage{times}
\usepackage{epsf}
\usepackage{graphicx}
\usepackage{epsfig}
\usepackage{epstopdf}
\usepackage{dcolumn}
\usepackage{bm}
\usepackage{float}
\usepackage{url}
\usepackage{CJK}
\usepackage{ulem}
\usepackage{makecell}
\usepackage{newtxtext, newtxmath}
\usepackage{multirow}
\usepackage{subfigure}
\usepackage{siunitx}
\usepackage[colorlinks=true,linkcolor=blue,citecolor=blue,urlcolor=blue]{hyperref}

\def\<{\langle}
\def\>{\rangle}
\def\({\left(}
\def\){\right)}
\def\[{\left[}
\def\]{\right]}

\def\sin{\mathop{\mathrm{sin}}\nolimits}
\def\cos{\mathop{\mathrm{cos}}\nolimits}

\usepackage{anyfontsize}
\begin{document}
\title{Finite-temperature Green's function theory of terahertz-induced phonon angular momentum in
polar crystals}
\author{Hong Sun}
\affiliation{Ministry of Education Key Laboratory of NSLSCS, Phonon Engineering Research Center of Jiangsu Province, Center for Quantum Transport and Thermal Energy Science, Institute of Physics Frontiers and Interdisciplinary Sciences, School of Physics and Technology, Nanjing Normal University, Nanjing 210023, China}

\author{Xiaozhe Li} 
\affiliation{Ministry of Education Key Laboratory of NSLSCS, Phonon Engineering Research Center of Jiangsu Province, Center for Quantum Transport and Thermal Energy Science, Institute of Physics Frontiers and Interdisciplinary Sciences, School of Physics and Technology, Nanjing Normal University, Nanjing 210023, China}

\author{Lifa Zhang}
\email{phyzlf@njnu.edu.cn}
\affiliation{Ministry of Education Key Laboratory of NSLSCS, Phonon Engineering Research Center of Jiangsu Province, Center for Quantum Transport and Thermal Energy Science, Institute of Physics Frontiers and Interdisciplinary Sciences, School of Physics and Technology, Nanjing Normal University, Nanjing 210023, China}
\date{\today}

\begin{abstract}
We develop a self-energy-dressed Green's-function framework for terahertz-induced phonon angular momentum in polar crystals. The rectified angular-momentum response is formulated as a second-order response and, within a dressed-bubble approximation, is expressed in terms of retarded phonon propagators weighted by mode-resolved polarization and angular-momentum matrix elements. Anharmonic self-energies enter through the dressed propagators, incorporating finite-temperature frequency renormalization and linewidth broadening directly into the response kernel. Applying this framework to wurtzite GaN, we show that the terahertz propagation direction and polarization select distinct rotational phonon channels: a nondegenerate $\mathrm{E}_{1}(\mathrm{TO})$--$\mathrm{A}_{1}(\mathrm{TO})$ channel with phase- and frequency-tunable angular momentum, and a degenerate $\mathrm{E}_{1}(\mathrm{TO})$ channel with helicity-selected response. Anharmonic broadening suppresses and smooths the resonant structures while preserving the characteristic phase dependence of each channel. An order-of-magnitude estimate based on the phonon inverse Faraday-effect framework gives an mT-scale effective magnetic field for representative electron--phonon coupling strengths. This work places coherent terahertz-driven circular-phonon physics in a first-principles-based response framework and provides a starting point for going beyond constant-damping driven-mode descriptions.
\end{abstract}
\makeatletter
\newcommand{\rmnum}[1]{\romannumeral #1}
\newcommand{\Rmnum}[1]{\expandafter\@slowromancap\romannumeral #1@}
\makeatother
\maketitle
\section{Introduction}
Intense terahertz (THz) excitation provides a powerful route for nonequilibrium control in condensed-matter systems, because low-energy collective modes can be coherently driven to large amplitudes by the electric-field component of a THz pulse~\cite{forst2011nonlinear,kampfrath2013resonant,Nicoletti:16}. This capability has been demonstrated in studies of transient insulator-to-metal transitions~\cite{rini2007control,PhysRevLett.108.136801,giorgianni2019overcoming}, ultrafast manipulation of magnetic order~\cite{forst2015spatially,afanasiev2021ultrafast,ilyas2024terahertz}, and light-enhanced superconducting correlations~\cite{rowe2023resonant,fava2024magnetic,eckhardt2024theory}, where coherent driving is used to reshape electronic, magnetic, or structural states. Recent developments in chiral and axial phonon physics~\cite{PhysRevLett.112.085503,PhysRevLett.115.115502,zhuexp,juraschek2025chiral} have added a further perspective to this route. THz and mid-infrared phonon driving are no longer viewed only as means of producing large-amplitude lattice displacements or rectified structural distortions; they also provide access to the phase structure and rotational character of lattice motion.
\par
In this context, the relative phase between driven vibrational components plays a central role~\cite{nova2017effective}. When two infrared-active phonon modes are coherently excited with a finite phase difference, the ions can move along circular or elliptical trajectories around their equilibrium positions~\cite{fwr7-6jj6}, thereby carrying mechanical phonon angular momentum~\cite{PhysRevLett.112.085503}. In ionic crystals, such rotational motion is accompanied by a phonon orbital magnetic moment~\cite{PhysRevMaterials.3.064405} and can generate magnetization through dynamical multiferroicity~\cite{PhysRevMaterials.1.014401}, where a time-dependent electric polarization produces a magnetic response~\cite{PhysRevLett.131.196801,7lm1-wm3y}. When coupled to electronic spin and orbital degrees of freedom, this rotational lattice motion can further give rise to light-induced magnetization~\cite{PhysRevResearch.4.013129, doi:10.1126/science.adi9601,PhysRevLett.132.096702,doi:10.1126/sciadv.ado0722} and angular-momentum transfer~\cite{PhysRevResearch.2.023275,choi2024real,minakova2026observation}, as discussed in terahertz-driven dynamical multiferroicity in SrTiO$_3$~\cite{basini2024terahertz}, phonon inverse Faraday effects~\cite{PhysRevLett.133.266702}, and related multiferroic responses~\cite{PhysRevResearch.3.L022011}.
\par
Despite these advances, a predictive quantum-response framework for THz-driven rotational lattice dynamics has not yet been fully established. Many existing studies describe the coherent dynamics of infrared-active phonons using an effective Lorentz-oscillator or mode-coordinate equation~\cite{PhysRevB.89.220301,9vb8-8gpc,mankowsky2014nonlinear}. In such descriptions, an infrared-active mode coordinate $\mathcal{Q}$ is driven by the electric field through its mode effective charge and obeys
\begin{equation}    
\ddot{\mathcal{Q}}+
\eta\dot{\mathcal{Q}}+
\frac{\partial}{\partial\mathcal{Q}}V(\mathcal{Q})
=
\sum_{i} Z_{i}\tilde{E}_{i}(t).
\end{equation}
Here $\eta$ is a phenomenological damping constant, $Z_{i}$ is the mode effective charge, $\tilde{E}_{i}$ is the $i$th component of the effective internal THz electric field acting on the lattice polarization, and $V(\mathcal{Q})$ denotes the harmonic or anharmonic lattice potential. Such phenomenological driven-mode descriptions have provided an intuitive and useful framework for resonant phonon driving, nonlinear phononics, and dynamical multiferroic responses~\cite{PhysRevMaterials.1.014401,PhysRevLett.118.054101,Caruso_2026}. However, their predictive power depends sensitively on the effective parameters entering the model. In particular, the damping constant, which controls both the decay and the phase lag of the driven phonon coordinate, is often introduced phenomenologically or taken from experimental linewidths. This becomes a nontrivial limitation for rotational lattice dynamics, because the resulting phonon angular momentum is determined not only by the amplitudes of the driven phonons but also by the relative phase between different vibrational components~\cite{PhysRevLett.112.085503}. A microscopic response formulation, in which THz mode selectivity, intermode angular-momentum matrix elements, frequency renormalization, and damping enter on the same footing, is therefore needed for a more predictive description of THz-induced phonon angular momentum.
\par
In this work, we develop a first-principles-based quantum-response framework for THz-induced phonon angular momentum in polar crystals. Starting from the dipole coupling between the lattice polarization and the effective internal THz electric field, we derive a second-order response kernel for the rectified, zero-output-frequency angular-momentum response. In the phonon eigenmode basis, the kernel is expressed in terms of mode-resolved polarization matrix elements, phonon angular-momentum matrix elements, and retarded phonon Green's functions. Finite-temperature anharmonic effects enter through phonon self-energies, which dress the retarded Green's functions and yield mode-dependent frequency renormalization and linewidth broadening.
\par
As a representative application, we study wurtzite GaN, an anisotropic polar crystal in which the infrared-active phonons and LO--TO splitting depend strongly on the propagation direction. This anisotropy allows different THz-driving geometries to selectively access the nondegenerate $\mathrm{E}_{1}(\mathrm{TO})$--$\mathrm{A}_{1}(\mathrm{TO})$ channel and the degenerate $\mathrm{E}_{1}(\mathrm{TO})$ pair, leading to distinct directions, phase dependences, and temperature dependences of the induced phonon angular momentum. Numerical results show that the nondegenerate channel produces a phase-dependent response shaped by inequivalent resonant tensor components and anharmonic broadening, including a frequency-tunable sign change, whereas the degenerate channel is governed by an antisymmetric field combination and displays a helicity-selected angular-momentum response. Estimates of the associated effective magnetic field further suggest that the induced angular momentum can be connected to magnetic signatures in an experimentally relevant range. Taken together, this theoretical formulation and its application to GaN bridge phenomenological driven-oscillator descriptions and first-principles finite-temperature response calculations of THz-controlled rotational lattice dynamics in polar crystals.

\section{Theory}

In this section, we derive the Green's-function kernel used below to evaluate THz-induced phonon angular momentum. The derivation connects the second-order dc response generated by the electric-dipole perturbation to a phonon-basis response kernel dressed by finite-temperature anharmonic self-energies. Detailed steps are provided in Apps.~\ref{ap:A} and \ref{ap:B}.

\subsection{Second-order time-averaged response to a THz electric field}\label{theory:A}

We first recast the THz-driven lattice dynamics in a nonlinear-response form. Rather than introducing a phenomenological damping term at the level of driven mode coordinates, we treat the electric field as a time-dependent perturbation coupled to the lattice polarization, $\hat{\mathcal H}^{\prime}(t)=-\sum_{\mu}\tilde{E}_{\mu}(t)\hat{P}_{\mu}$~\cite{RevModPhys.66.899}. Because this perturbation is linear in the ionic displacement, whereas the phonon angular momentum is bilinear in displacement and velocity, the induced angular momentum vanishes at linear order. The leading contribution is therefore a second-order response. Following the standard second-order Kubo expansion~\cite{doi:10.1143/JPSJ.12.570}, the induced angular momentum takes the causal form
\begin{equation}\label{eq:2ndresponse}
\begin{aligned}
{J}_{\mathrm{ph}}^{\alpha}(t) &= 
\lim_{s\to 0^{+}}\sum_{\mu\nu}\int_{0}^{+\infty}du\int_{0}^{+\infty}dv \Pi_{\mu\nu}^{\alpha}(u,v)\\
&\times e^{s(2t-2u-v)}\tilde{E}_{\mu}(t-u)\tilde{E}_{\nu}(t-u-v).
\end{aligned}
\end{equation}
Here ${J}_{\mathrm{ph}}^{\alpha}(t)$ is the time-dependent phonon angular momentum along the $\alpha$ direction, and $\mu,\nu$ label Cartesian components of the internal electric field. The quantity $\Pi_{\mu\nu}^{\alpha}(u,v)=(-i/\hbar)^2\langle[[\hat{J}_{I}^{\alpha}(u+v),\hat{P}^{I}_{\mu}(v)],\hat{P}^{I}_{\nu}(0)]\rangle$ is the retarded time-domain kernel relating the angular-momentum response to the lattice polarization, with the subscript $I$ denoting the interaction picture and $\langle\cdots\rangle$ the equilibrium average. The factor involving $s\to0^{+}$ arises from the adiabatic switching of the perturbation and fixes the causal prescription.
\par
Since we are interested in the time-averaged angular-momentum response generated by the THz pulse, we focus on the corresponding zero-output-frequency component of $J_{\mathrm{ph}}^{\alpha}(t)$. We first define the response kernel by Fourier transforming the retarded time-domain kernel with respect to the two positive relative times,
\begin{equation}\label{eq:chikernel_dc}
\begin{aligned}
\chi^{\alpha}_{\mu\nu}(\omega,-\omega)&
=\frac{1}{2\pi}\int_{0}^{+\infty}du\int_{0}^{+\infty}dv
\Pi^{\alpha}_{\mu\nu}(u,v)\\
&\times e^{i(0+i0^{+})u}e^{i(-\omega+i0^{+})v}.
\end{aligned}
\end{equation}
The infinitesimal factors above inherit the causal prescription fixed by the adiabatic switching in the time-domain response. The derivation of Eqs.~\eqref{eq:2ndresponse} and~\eqref{eq:chikernel_dc}, including the interaction-picture density-matrix expansion and the Fourier transformation leading to the zero-output-frequency response kernel, is given in App.~\ref{ap:A}. Within this framework, the zero-output-frequency component is obtained as
\begin{equation}\label{eq:Jph_dc}
\mathcal{J}^{\alpha}_{\mathrm{ph},0}
=
\sum_{\mu\nu}
\int_{-\infty}^{+\infty}d\omega\,
\chi_{\mu\nu}^{\alpha}(\omega,-\omega)
\tilde{E}_{\mu}(\omega)
\tilde{E}_{\nu}(-\omega).
\end{equation}
Here, $\chi_{\mu\nu}^{\alpha}(\omega,-\omega)$ contains the intrinsic lattice response that converts two field-induced polarization perturbations into a phonon-angular-momentum expectation value, whereas the product $\tilde{E}_{\mu}(\omega)\tilde{E}_{\nu}(-\omega)$ selects the zero-total-frequency component of the driving field. In the following, we evaluate $\chi_{\mu\nu}^{\alpha}(\omega,-\omega)$ in the phonon eigenmode basis, reducing the formal second-order kernel to a mode-resolved Green's-function expression weighted by polarization and angular-momentum matrix elements.

\subsection{Mode-resolved Green's-function representation of the response kernel}\label{theory:B}

Having obtained the response kernel in Eq.~\eqref{eq:chikernel_dc}, we now evaluate its operator structure in the phonon eigenmode basis. This representation expresses both the phonon-angular-momentum and polarization operators in terms of phonon normal modes, and provides the starting point for reducing the formal second-order kernel to a mode-resolved Green's-function expression. The resulting formulation separates two microscopic ingredients: the mode-resolved polarization matrix elements, which determine how the THz field couples to infrared-active phonons, and the off-diagonal phonon-angular-momentum matrix elements, which weight the corresponding intermode phonon-pair channels.
\par
To express the response kernel in a Green's-function form, we first introduce the standard displacement-like phonon field operator $\hat{\mathcal{A}}_{\mathbf{q}c}=\hat{a}_{\mathbf{q}c}+\hat{a}^{\dagger}_{-\mathbf{q}c}$ and the velocity-like operator 
$\hat{\mathcal{B}}_{\mathbf{q}c}=\hat{a}_{\mathbf{q}c}-\hat{a}^{\dagger}_{-\mathbf{q}c}$~\cite{mahan2000many}, 
where $\hat{a}_{\mathbf{q}c}$ and $\hat{a}^{\dagger}_{\mathbf{q}c}$ are the annihilation and creation operators of phonon mode $(\mathbf{q},c)$, respectively. Under harmonic time evolution, these operators satisfy $i\partial_{t}\hat{\mathcal{A}}^{I}_{\mathbf{q}c}(t)=\omega_{\mathbf{q}c}\hat{\mathcal{B}}^{I}_{\mathbf{q}c}(t)$, 
which allows correlators involving $\hat{\mathcal{B}}$ to be rewritten as time derivatives of $\hat{\mathcal{A}}$--$\hat{\mathcal{A}}$ correlators. In terms of these operators, the phonon angular momentum can be written as
\begin{equation}\label{eq5}
\hat{J}^{\alpha}_{I}(t)
=
\sum_{\mathbf{q},ab}
\mathcal{L}^{\alpha}_{\mathbf{q},ab}
\hat{\mathcal{A}}^{I}_{\mathbf{q}a}(t)
\hat{\mathcal{B}}^{I}_{-\mathbf{q}b}(t),
\end{equation}
where
\begin{equation}
\mathcal{L}^{\alpha}_{\mathbf{q},ab}
=
-\frac{i\hbar}{2}
\sqrt{\frac{\omega_{\mathbf{q}b}}{\omega_{\mathbf{q}a}}}
\sum_{\kappa}
\left(
\mathbf{e}_{\mathbf{q}a,\kappa}
\times
\mathbf{e}^{*}_{\mathbf{q}b,\kappa}
\right)_{\alpha}.
\end{equation}
Here $\mathbf{e}_{\mathbf{q}a,\kappa}$ is the component of the phonon eigenvector associated with atom $\kappa$ for mode $(\mathbf{q},a)$ with frequency $\omega_{\mathbf{q}a}$. The matrix element $\mathcal{L}^{\alpha}_{\mathbf{q},ab}$ encodes the angular-momentum structure of a phonon-mode pair $(a,b)$ at the same wave vector $\mathbf{q}$.
\par
Likewise, the lattice polarization is expanded in the same phonon-field basis as
\begin{equation}\label{eq7}
\hat{P}_{\mu}^{I}(t)
=
\sum_{a}
\mathcal{P}_{a}^{\mu}{\hat{\mathcal{A}}}^{I}_{a}(t),
\end{equation}
where $\mathcal{P}_{a}^{\mu}$ is the mode-resolved polarization matrix element. Within the dipole approximation, it is given by
\begin{equation}
\mathcal{P}_{a}^{\mu}=\sum_{\mathbf{q}}\sum_{\kappa\nu}
\sqrt{\frac{\hbar N}{2m_{\kappa}\omega_{\mathbf{q}a}}}
\mathcal{Z}^{*}_{\kappa,\mu\nu} e^{\nu}_{\mathbf{q}a,\kappa}\delta_{\mathbf{q},0}.
\end{equation}
Here $N$ is the number of unit cells, $m_{\kappa}$ is the mass of atom $\kappa$, and $\mathcal{Z}^{*}_{\kappa,\mu\nu}$ is the corresponding Born effective charge tensor~\cite{PhysRevB.55.10355}. The factor $\delta_{\mathbf{q},0}$ reflects the dipole selection rule: a spatially uniform electric field couples only to zone-center phonon modes with nonzero polarization matrix elements, i.e., infrared-active modes.
\par
Substituting the operator expressions in Eqs.~\eqref{eq5} and \eqref{eq7} into the time-domain kernel $\Pi_{\mu\nu}^{\alpha}(u,v)$ generates four-operator phonon correlation functions. Their explicit expansion and bubble-level reduction are given in App.~\ref{ap:B1}. Analogous to the decoupling commonly used in Green--Kubo and double-time Green's-function treatments of phonon transport~\cite{PhysRev.139.A1569,PhysRevB.5.3909,PhysRevB.106.024312}, we reduce these four-point functions to products of two-point correlators by retaining the cross-vertex contractions. For a generic correlator
$\langle\hat{X}_{1}(u+v)\hat{X}_{2}(u+v)\hat{X}_{3}(v)\hat{X}_{4}(0)\rangle$, these contractions are $\langle\hat{X}_{1}\hat{X}_{3}\rangle\langle\hat{X}_{2}\hat{X}_{4}\rangle+\langle\hat{X}_{1}\hat{X}_{4}\rangle\langle\hat{X}_{2}\hat{X}_{3}\rangle$, which connect the angular-momentum vertex to the two polarization vertices. The internal pairing $\langle\hat{X}_{1}(u+v)\hat{X}_{2}(u+v)\rangle\langle\hat{X}_{3}(v)\hat{X}_{4}(0)\rangle$ is not retained, because it does not describe two field-induced propagation lines connecting the polarization vertices to the angular-momentum vertex.
\par
This reduction selects the bubble contribution to the nonlinear response. It retains the part of the induced phonon angular momentum that is constructed from two field-driven single-phonon propagation processes. Equivalently, by writing $\hat{\mathbf{u}} = \langle\hat{\mathbf{u}}\rangle+\delta\hat{\mathbf{u}}$, the retained contribution corresponds schematically to the coherent mean-displacement term $\langle \hat{\mathbf{u}}\rangle^{(1)}\times \langle \hat{\dot{\mathbf{u}}}\rangle^{(1)}$, with the linear phonon amplitudes dressed by the propagators. Contributions beyond this bubble-level treatment would involve field-induced correlations in the fluctuation sector, including vertex-corrected and covariance-type terms such as changes in $\langle \delta\hat{\mathbf{u}}\times\delta\hat{\dot{\mathbf{u}}} \rangle^{(2)}$~\cite{PhysRevB.106.024312,PhysRevB.107.054311}. Such terms are expected to become more important when the drive strongly modifies lattice fluctuations or phonon-pair correlations, as occurs in driven fluctuation quenching, parametric phonon amplification, and squeezed-phonon generation, or when anharmonic mode mixing and phonon coherences invalidate an independent-mode propagation picture~\cite{trigo2013fourier,doi:10.1073/pnas.1809725115,fechner2024quenched}. In the weak-driving, near-equilibrium quasiparticle regime considered here, we neglect these beyond-bubble corrections and incorporate the main finite-temperature effects through dressed single-phonon propagators.
\par
With these approximations, the time-domain kernel is reduced to
\begin{equation}\label{eq:Pi_greenfunction}
\begin{aligned}
\Pi_{\mu\nu}^{\alpha}(u,v)
&= i \sum_{abcd}
\tilde{\mathcal{L}}^{\alpha}_{ab}\,
\mathcal{P}_{c}^{\mu}\mathcal{P}_{d}^{\nu}
\Big[
G^{R}_{ac}(u)\,
\partial_{u+v} G^{R}_{bd}(u+v)
\\
&\qquad\qquad
+\, G^{R}_{ad}(u+v)\,
\partial_{u} G^{R}_{bc}(u)
\Big],
\end{aligned}
\end{equation}
where $\tilde{\mathcal{L}}_{ab}^{\alpha}\equiv\mathcal{L}^{\alpha}_{ab}/\omega_{b}$, and $G^{R}$ is the retarded phonon Green's function defined by~\cite{PhysRev.128.2589}
\begin{equation}
G^{R}_{ab}(t)=
\(-\frac{i}{\hbar}\)\theta(t)
\left<
\[
\hat{\mathcal{A}}^{I}_{a}(t),\hat{\mathcal{A}}^{I,\dagger}_{b}(0)
\]
\right>.
\end{equation}
The time derivatives here arise from the velocity-like operator $\hat{\mathcal{B}}$ in the phonon angular momentum. The wave-vector index has been omitted in this propagator form because translational symmetry makes the phonon propagators diagonal in crystal momentum, while the two polarization vertices restrict the field-coupled modes to $\mathbf{q}=0$. Therefore, the response is evaluated within the zone-center sector: the polarization matrix elements $\mathcal{P}_{a}^{\mu}$ select the infrared-active phonons driven by the THz field, whereas the angular-momentum matrix elements $\mathcal{L}^{\alpha}_{ab}$ convert the field-driven mode-pair response into angular momentum. For the time-reversal-symmetric lattice considered here, the zone-center phonon eigenvectors can be chosen real, so the diagonal elements vanish, $\mathcal{L}^{\alpha}_{aa}=0$. The induced phonon angular momentum is thus carried by intermode processes.
\par
Substituting Eq.~\eqref{eq:Pi_greenfunction} into Eq.~\eqref{eq:chikernel_dc} and using the Fourier representation of $G^{R}$, we obtain the response kernel as
\begin{widetext}
\begin{equation}\label{eq:chikernel_gr}
        \chi_{\mu\nu}^{\alpha}(\omega,-\omega)=
        -\sum_{abcd}
        \frac{\tilde{\mathcal{L}}^{\alpha}_{ab}\mathcal{P}^{\mu}_{c}\mathcal{P}^{\nu}_{d}}{(2\pi)^{3}}
        \int_{-\infty}^{+\infty}d\omega_{1}
        \int_{-\infty}^{+\infty}d\omega_{2}
\frac{
\omega_{2}G^{R}_{ac}(\omega_1)G^{R}_{bd}(\omega_2)
+
\omega_{1}G^{R}_{bc}(\omega_1)G^{R}_{ad}(\omega_2)
}{
\left(\omega_1+\omega_2-i0^{+}\right)
\left(\omega+\omega_2-i0^{+}\right)
}.
\end{equation}
\end{widetext}
The algebraic reduction from the four-operator kernel to the Green's-function convolution in Eq.~\eqref{eq:chikernel_gr} is provided in App.~\ref{ap:B1}. In this expression, the kernel is organized as a convolution of two retarded phonon propagators. Its magnitude is controlled jointly by the matrix-element prefactors and the frequency dependence of the two propagators. The causal denominators couple the two internal frequencies and determine how different spectral components contribute to the zero-output-frequency response. This form makes it possible to incorporate finite-temperature self-energy corrections directly through the dressed phonon Green's functions, as discussed in the next subsection.

\subsection{Self-energy dressing and broadened response kernel}\label{theory:C}

Since the self-energy dressing of retarded phonon Green's functions at finite temperature is standard in anharmonic lattice dynamics~\cite{PhysRev.128.2589,Cowley_1968}, we do not repeat its derivation here. Instead, we use the dressed Green's function to rewrite the response kernel in terms of renormalized phonon frequencies and damping parameters, thereby making the mode-pair structure of the broadened response explicit. For the well-defined zone-center optical modes considered in this work, we retain only the diagonal component of the self-energy in the phonon eigenmode basis. The retarded Green's function is therefore approximated as $G_{ab}^{R}(\omega)\simeq\delta_{ab}G_{a}^{R}(\omega)$. In the harmonic eigenmode basis, the off-diagonal propagators vanish at zeroth order and are generated only through off-diagonal self-energy insertions in the Dyson equation. They are thus neglected within the present diagonal quasiparticle approximation.
\par
The resulting diagonal retarded Green's function is written as
\begin{equation}
G^{R}_{a}(\omega)=\frac{1}{\hbar}\frac{2\omega_{a}}
    {
    (\omega+i0^{+})^2-
    \omega_{a}^2-
    2\omega_{a}\Sigma^{R}_{a}(\omega)
    },
\end{equation}
where $\Sigma^{R}_{a}(\omega)=\Delta_{a}(\omega)-i\Gamma_{a}(\omega)$ is the mode-resolved retarded anharmonic self-energy. Its real part gives the frequency renormalization, whereas its imaginary part describes anharmonic damping and spectral broadening. To obtain a compact working expression, we use the quasiparticle form of the dressed Green's function by evaluating the self-energy near the phonon pole.  Defining $\Delta_{a}\equiv \Delta_{a}(\omega_{a})$, $\Gamma_{a}\equiv \Gamma_{a}(\omega_{a})$, and $\tilde{\omega}_{a}=\omega_{a}+\Delta_{a}$, we obtain
\begin{equation}\label{eq13}
 G^{R}_{a}(\omega)\simeq\frac{1}{\hbar}\(
\frac{1}{\omega-\tilde{\omega}_{a}+i\Gamma_{a}}-
\frac{1}{\omega+\tilde{\omega}_{a}+i\Gamma_{a}}\).
\end{equation}
In this pole representation, $\Gamma_a$ corresponds to the damping half-width of the quasiparticle resonance. Thus, finite-temperature anharmonic effects are incorporated through shifted phonon resonances and finite broadening.
\par
Substituting this form into the convolution expression for the response kernel in Eq.~\eqref{eq:chikernel_gr}, and evaluating the internal frequency integrals by the residue theorem as detailed in App.~\ref{ap:B2}, gives
\begin{equation}\label{eq14}
\chi^{\alpha}_{\mu\nu}(\omega,-\omega)=\frac{1}{2\pi\hbar^2}\sum_{ab}
    \tilde{\mathcal{L}}^{\alpha}_{ab}
    \mathcal{K}_{ab}^{\mu\nu}(\omega),
\end{equation}
where 
\begin{equation}\label{eq15}
\mathcal{K}_{ab}^{\mu\nu}(\omega)=
    \sum_{\eta_{a},\eta_{b}=\pm 1}
    \frac{\eta_{a}\eta_{b}\zeta_{b}^{\eta_{b}}}{\zeta_{a}^{\eta_{a}}+\zeta_{b}^{\eta_{b}}}
    \(
\frac{
\mathcal{P}_{a}^{\mu}
\mathcal{P}_{b}^{\nu}
}
{\omega+\zeta^{\eta_{b}}_{b}}
+
\frac{
\mathcal{P}_{b}^{\mu}
\mathcal{P}_{a}^{\nu}
}
{\omega+\zeta^{\eta_{a}}_{a}}
    \).
\end{equation}
Here $\zeta^{\eta_{a}}_{a}=\eta_{a}\tilde{\omega}_{a}-i\Gamma_{a}$ denotes the positive- or negative-frequency complex pole of the dressed retarded Green's function. Eqs.~\eqref{eq14} and \eqref{eq15} therefore show that the response is resolved into four pole channels labeled by $\eta_{a},\eta_{b}=\pm 1$. The renormalized frequencies determine the shifted resonant positions, whereas the damping half-widths control the broadening and overlap of the corresponding mode-pair contributions.
\par
With increasing temperature, anharmonic scattering usually enhances the damping half-widths, which broaden the resonant denominators and suppress sharply resolved peak structures. At the same time, the broadened poles extend the frequency range over which a given mode pair contributes, thereby modifying the relative weights of different channels in the total response. The resulting kernel also satisfies the conjugation relation $\chi_{\mu\nu}^{\alpha}(\omega,-\omega)=\left[\chi_{\mu\nu}^{\alpha}(-\omega,\omega)\right]^{*}$, which follows from the conjugation properties of the dressed poles and the angular-momentum matrix elements. This relation ensures that the induced dc phonon angular momentum is real for a physical driving field. The proof of this property, together with the narrow-linewidth limit of Eq.~\eqref{eq15}, is given in App.~\ref{ap:B3}.

\section{Results and Discussion}\label{Results}
We now apply the self-energy-dressed response kernel to wurtzite GaN. GaN is a useful test system for the present purpose because its low-energy THz response is governed by infrared-active lattice polarization, making it directly compatible with the dipole-coupling formulation developed above. In addition, the uniaxial wurtzite structure gives rise to direction-dependent LO--TO splitting and anisotropic infrared-active phonons~\cite{PhysRevB.7.743}, so that different transverse-field geometries select different pairs of zone-center modes. The relevant infrared-active TO sector is also relatively simple, consisting of the nondegenerate $\mathrm{A}_{1}(\mathrm{TO})$ mode and the doubly degenerate $\mathrm{E}_{1}(\mathrm{TO})$ modes~\cite{PhysRevLett.86.906}. These features make GaN a minimal platform for comparing nondegenerate and degenerate THz-driven rotational phonon responses within the same microscopic framework. The first-principles computational details, convergence tests, and extraction of finite-temperature spectral parameters are provided in the Supplemental Material~\cite{SM} (see also references~\cite{PhysRevB1996,
PhysRevB1994,
PhysRevLett1996,
PhysRevB.61.6720,
exp,
PhysRevB.56.R10024,
knoop2024tdep,
PhysRevB.84.180301,
PhysRevB.87.104111} therein).
\par
\begin{figure}[htbp]
\centering
\includegraphics[width=1\columnwidth]{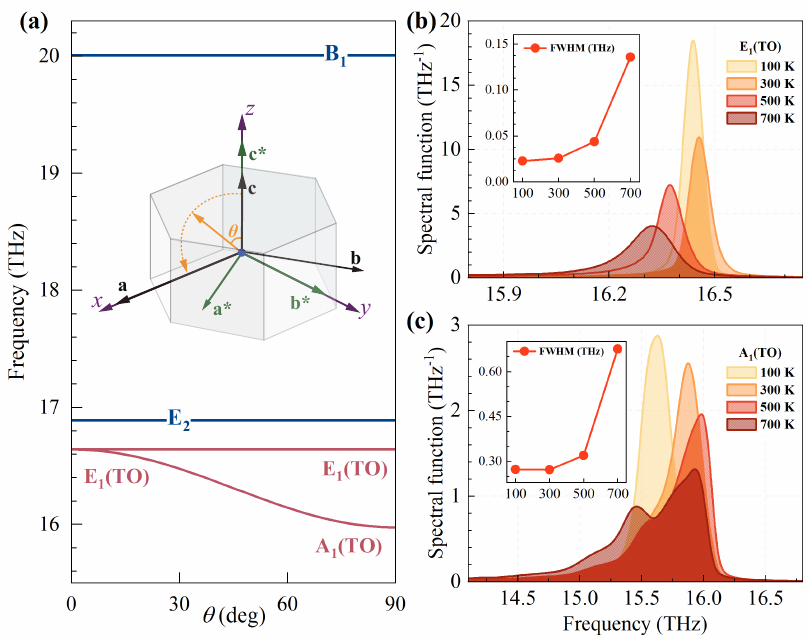}
\caption{Direction-dependent infrared-active TO modes and finite-temperature spectral functions in wurtzite GaN. (a) Zone-center optical-phonon frequencies in the long-wavelength limit as the phonon wave-vector direction is rotated from the $c$ axis to the in-plane $x$ direction. The inset defines the rotation angle $\theta$ and the crystallographic axes. (b),(c) Temperature-dependent spectral functions of the selected $\mathrm{E}_{1}(\mathrm{TO})$ and $\mathrm{A}_{1}(\mathrm{TO})$ modes, respectively. The insets show the corresponding full widths at half maximum (FWHM).}
\label{fig1}
\end{figure}
Fig.~\ref{fig1} summarizes the direction-dependent zone-center infrared-active TO modes and their finite-temperature spectral properties. Owing to the nonanalytic long-range dipole--dipole interaction, the long-wavelength optical phonon frequencies depend on the direction from which the $\Gamma$ point is approached. Fig.~\ref{fig1}(a) shows this directional variation of the infrared-active TO modes, which, together with the transverse nature of the THz electric field, determines the mode pairs driven in different propagation geometries. For propagation along the $c$ axis, the transverse THz field lies in the basal plane and couples to the two degenerate $\mathrm{E}_{1}(\mathrm{TO})$ components. Their phase-coherent superposition can generate circular or elliptical ionic motion in the $xy$ plane, and therefore a phonon angular momentum along the $z$ direction. By contrast, for propagation along an in-plane direction, the transverse field may contain both in-plane and out-of-plane components, which couple respectively to the $\mathrm{E}_{1}(\mathrm{TO})$ and $\mathrm{A}_{1}(\mathrm{TO})$ modes. The selected nondegenerate mode pair can then produce coherent ionic motion in the $yz$ plane and generate an $x$-directed phonon angular momentum. Thus, the directional infrared response of wurtzite GaN naturally separates the THz-induced angular-momentum response into two representative channels: the degenerate $\mathrm{E}_{1}(\mathrm{TO})$ pair and the nondegenerate $\mathrm{E}_{1}(\mathrm{TO})$--$\mathrm{A}_{1}(\mathrm{TO})$ pair.
\par
The finite-temperature spectral functions of these TO modes are shown in Figs.~\ref{fig1}(b) and \ref{fig1}(c). For the $\mathrm{E}_{1}(\mathrm{TO})$ mode, the spectral function remains dominated by a single well-defined peak over the temperature range considered. The peak position changes only modestly with increasing temperature, whereas the linewidth increases continuously, as quantified by the FWHM shown in the inset of Fig.~\ref{fig1}(b). This behavior indicates that anharmonic damping, rather than frequency renormalization, provides the dominant finite-temperature correction for this mode. The $\mathrm{A}_{1}(\mathrm{TO})$ mode exhibits stronger anharmonic broadening. As shown in Fig.~\ref{fig1}(c), it remains quasiparticle-like from 100 to 500 K, but the spectrum at 700 K develops a two-peak structure. In this regime, a description in terms of a single renormalized frequency and a single linewidth becomes ambiguous, and the diagonal quasiparticle approximation used in the response kernel is no longer quantitatively reliable for this mode. Therefore, in the following temperature-dependent analysis of the nondegenerate channel, we focus on the 100--500 K range, where both selected TO modes can still be parameterized by well-defined peak positions and FWHM values.
\par
\begin{figure}[htbp]
\centering
\includegraphics[width=1\linewidth]{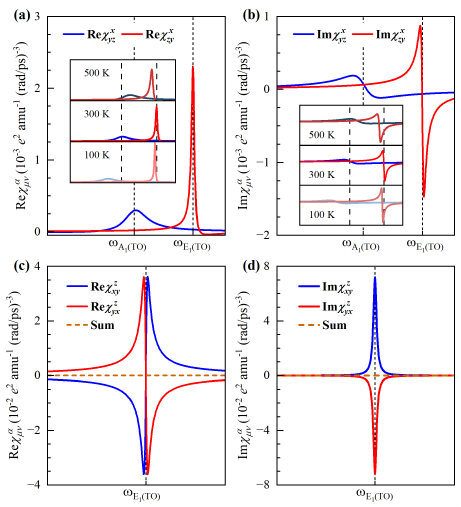}
\caption{Frequency-resolved response kernels in wurtzite GaN. (a),(b) Real and imaginary parts of the response kernel $\chi_{yz}^{x}$ and $\chi_{zy}^{x}$ for the nondegenerate $\mathrm{E}_{1}(\mathrm{TO})$--$\mathrm{A}_{1}(\mathrm{TO})$ channel selected by an in-plane propagation geometry. The insets show the corresponding temperature evolution. (c),(d) Real and imaginary parts of $\chi_{xy}^{z}$ and $\chi_{yx}^{z}$ for the degenerate $\mathrm{E}_{1}(\mathrm{TO})$ pair selected by propagation along the $c$ axis; the dashed curves denote the direct sum of the two tensor components. In all panels, vertical dashed lines mark the relevant TO-mode frequencies, and the main curves are evaluated at 300 K.}
\label{fig2}
\end{figure}
\begin{figure*}[htbp]
\centering
\includegraphics[width=0.99\linewidth]{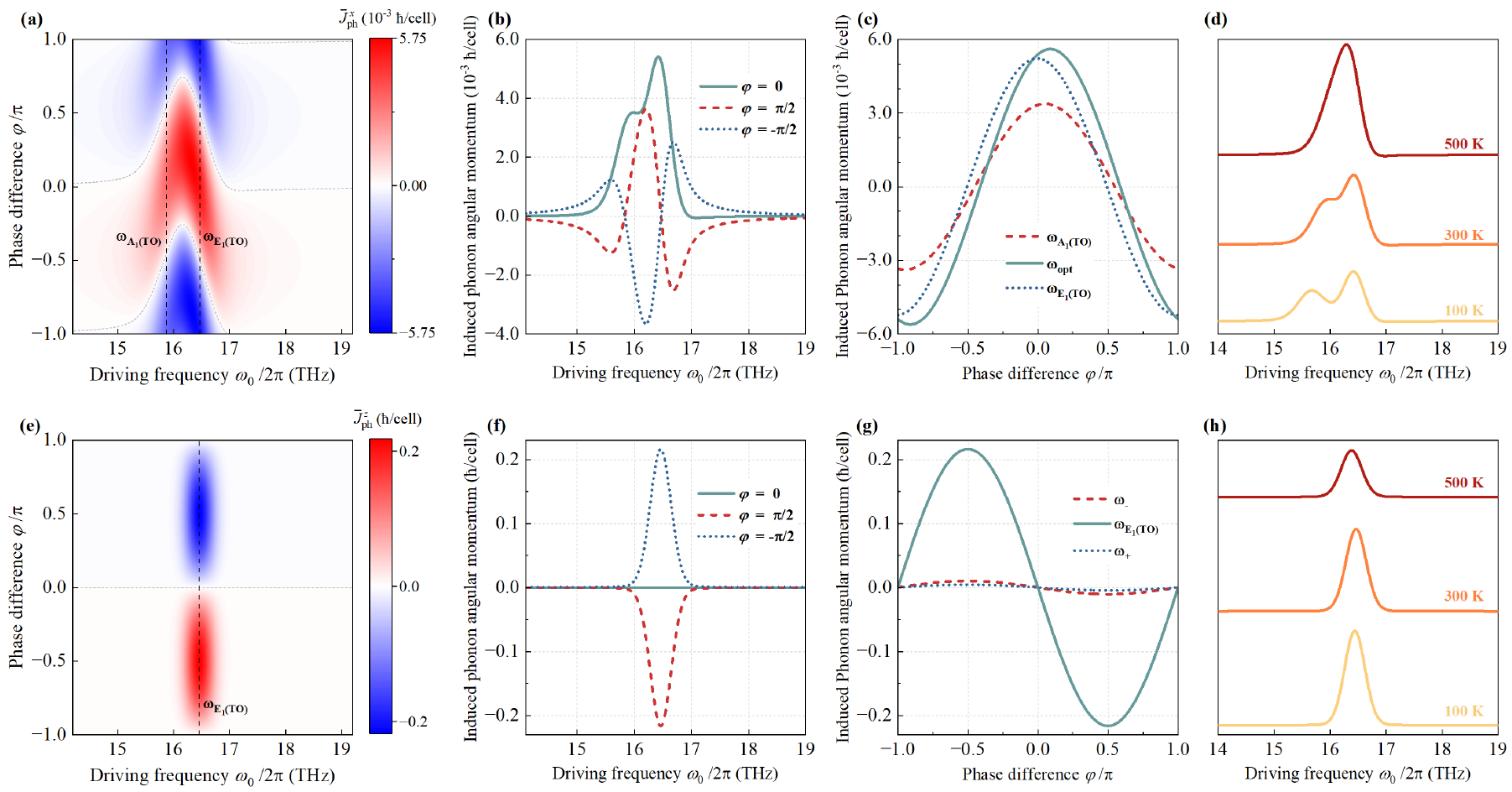}
\caption{Pulse-averaged phonon angular momentum per unit cell induced by a Gaussian-enveloped THz pulse. (a)--(d) Results for the nondegenerate $\mathrm{E}_{1}(\mathrm{TO})$--$\mathrm{A}_{1}(\mathrm{TO})$ channel, giving the $x$ component of the pulse-averaged phonon angular momentum. (e)--(h) Corresponding results for the degenerate $\mathrm{E}_{1}(\mathrm{TO})$ pair selected by propagation along the $c$ axis, giving the $z$ component. (a),(e) Dependence on the central driving frequency, plotted as $\omega_{0}/2\pi$, and on the phase difference $\varphi$. (b),(f) Frequency dependence for representative phase differences corresponding to linearly and circularly polarized THz fields. (c),(g) Phase dependence at representative driving frequencies. (d),(h) Temperature evolution of the frequency-dependent pulse-averaged angular momentum. Curves are vertically offset for clarity. Except for the temperature-evolution panels in (d) and (h), all results are evaluated at 300 K. The pulse parameters are $\tilde{E}_{0}=0.1$ MV/cm, $t_{0}=0$, and a temporal FWHM of 1.5 ps, corresponding to $\sigma = \mathrm{FWHM}/[2\sqrt{2\ln 2}]$. Vertical dashed lines in (a) and (e) mark the relevant TO-mode frequencies. Scale factors for the angular momentum are indicated on the axes and color bars.}
\label{fig3}
\end{figure*}
\par
The calculated response kernels $\chi^{\alpha}_{\mu\nu}(\omega,-\omega)$ at 300 K are shown in Fig.~\ref{fig2}. We first consider the nondegenerate $\mathrm{E}_{1}(\mathrm{TO})$--$\mathrm{A}_{1}(\mathrm{TO})$ channel, for which the relevant tensor components are $\chi_{yz}^{x}$ and $\chi_{zy}^{x}$. As shown in Figs.~\ref{fig2}(a) and \ref{fig2}(b), the real parts of both components exhibit resonant enhancements near the $\mathrm{A}_{1}(\mathrm{TO})$ and $\mathrm{E}_{1}(\mathrm{TO})$ frequencies, whereas the imaginary parts display sign-changing dispersive structures around the same characteristic frequencies. The different line shapes of $\chi_{yz}^{x}$ and $\chi_{zy}^{x}$ reflect the tensorial nature of the second-order response: exchanging the two transverse field components changes the ordering of the polarization vertices and the associated mode-pair factors, so the two components need not contribute symmetrically. The insets further show that this inequivalence persists with increasing temperature, while anharmonic self-energy broadening reduces the peak amplitudes and smooths the frequency-dependent structures, consistent with the increase in the FWHM discussed in Fig.~\ref{fig1}.
\par
By contrast, the degenerate in-plane $\mathrm{E}_{1}(\mathrm{TO})$ pair exhibits an antisymmetric relation between $\chi_{xy}^{z}$ and $\chi_{yx}^{z}$, as shown in Figs.~\ref{fig2}(c) and \ref{fig2}(d). For both the real and imaginary parts, the two components have equal magnitudes and opposite signs within numerical accuracy, so their direct sum vanishes, as indicated by the dashed curves. This cancellation is the expected consequence of the rotational symmetry of the degenerate in-plane channel, confirming that the microscopic response kernel derived here respects the required symmetry constraint~\cite{Birss1964}. It should be emphasized, however, that the cancellation of the direct sum does not imply that the degenerate channel is inactive. Rather, it indicates that this channel couples to the antisymmetric, phase-sensitive field product. After contraction with the THz field in Eq.~\eqref{eq:Jph_dc}, the relevant contribution is governed by $\tilde{E}_x(\omega)\tilde{E}_y(-\omega)-\tilde{E}_y(\omega)\tilde{E}_x(-\omega)$ and is therefore sensitive to the handedness of the in-plane driving field. Thus, circularly or elliptically polarized THz excitation can generate a sizable $\mathcal{J}_{\mathrm{ph,0}}^{z}$ through the degenerate in-plane channel.
\par
To further clarify how the two response-kernel structures identified above affect the induced phonon angular momentum, we evaluate the pulse-averaged response driven by a Gaussian-enveloped THz pulse with two orthogonal transverse electric-field components,
\begin{subequations}
\begin{align}
\tilde{E}_{\mu}(t)&=\tilde{E}_{0}g(\tau)\cos(\omega_{0}\tau),\\
\tilde{E}_{\nu}(t)&=\tilde{E}_{0}g(\tau)\cos(\omega_{0}\tau+\varphi).
\end{align}
\end{subequations} 
Here $\tau=t-t_{0}$ is the time measured from the pulse center, $\tilde{E}_{0}$ is the effective internal field amplitude, $g(\tau)=\exp[-\tau^{2}/(2\sigma^{2})]$ is the Gaussian temporal envelope with width $\sigma$, $\omega_{0}$ is the central driving angular frequency, and $\varphi$ is the phase difference between the two transverse field components. The numerical implementation of the Gaussian-pulse response and the effective-field estimate are summarized in App.~\ref{ap:C}.
\par
For a finite pulse, the quantity $\mathcal{J}_{\mathrm{ph},0}^{\alpha}$ obtained from Eq.~\eqref{eq:Jph_dc} strictly represents the zero-output-frequency Fourier weight of the transient rectified response, rather than a steady-state dc angular momentum. Following the finite-pulse normalization discussed in App.~\ref{ap:C1}, we therefore report the pulse-averaged angular momentum, defined as $\overline{J}_{\mathrm{ph}}^{\alpha}=\mathcal{J}_{\mathrm{ph},0}^{\alpha}/\tau_{\mathrm{av}}$. In the numerical results below, we choose $\tau_{\mathrm{av}}=2\tau_{\mathrm{FWHM}}$, where $\tau_{\mathrm{FWHM}}$ is the FWHM of the Gaussian field envelope. This choice fixes the normalization of the pulse-averaged quantity; changing $\tau_{\mathrm{av}}$ would rescale the reported magnitude but would not affect the frequency, phase, or temperature dependences determined by the response kernel and the pulse spectrum.
\par
The resulting pulse-averaged phonon angular momenta are presented in Fig.~\ref{fig3}. The upper panels [Figs.~\ref{fig3}(a)--\ref{fig3}(d)] show $\overline{J}_{\mathrm{ph}}^{x}$ for the nondegenerate $\mathrm{E}_{1}(\mathrm{TO})$--$\mathrm{A}_{1}(\mathrm{TO})$ channel, whereas the lower panels [Figs.~\ref{fig3}(e)--\ref{fig3}(h)] show $\overline{J}_{\mathrm{ph}}^{z}$ for the degenerate $\mathrm{E}_{1}(\mathrm{TO})$ channel. Fig.~\ref{fig3}(a) shows the frequency--phase map of $\overline{J}_{\mathrm{ph}}^{x}$, which visualizes its joint dependence on the central driving frequency $\omega_{0}$ and the phase difference $\varphi$. The response is mainly concentrated in the frequency window between the two TO resonances, and its sign and magnitude are strongly modulated by $\varphi$. Notably, the maximum response does not occur exactly at either $\omega_{\mathrm{A}_{1}(\mathrm{TO})}$ or $\omega_{\mathrm{E}_{1}(\mathrm{TO})}$, but is shifted to an intermediate driving frequency. This behavior is consistent with the inequivalent frequency structures of $\chi_{yz}^{x}$ and $\chi_{zy}^{x}$ discussed in Fig.~\ref{fig2}, together with their contraction with the finite-bandwidth THz pulse.
\par
Fig.~\ref{fig3}(b) further extracts the frequency dependence for three representative polarization states: left-circularly polarized ($\varphi=-\pi/2$), linearly polarized ($\varphi=0$), and right-circularly polarized ($\varphi=\pi/2$) THz fields. The finite response under linear polarization shows that the nondegenerate $\mathrm{E}_{1}(\mathrm{TO})$--$\mathrm{A}_{1}(\mathrm{TO})$ channel can generate angular momentum through the intrinsic phase lag between the driven modes. Circularly polarized fields further tune both the magnitude and the sign of $\overline{J}_{\mathrm{ph}}^{x}$ through the handedness of the drive. In addition, for a fixed polarization state, the induced angular momentum can change sign as the driving frequency is varied, providing a direct signature of the underlying frequency-dependent response kernel. Fig.~\ref{fig3}(c) resolves the phase dependence of $\overline{J}_{\mathrm{ph}}^{x}$ at representative driving frequencies. The response is approximately sinusoidal in $\varphi$, with an amplitude and phase offset that depend on $\omega_{0}$. This behavior is consistent with the Lorentz-oscillator picture that the relative phase controls the rotational component of the driven lattice motion, while the present calculation determines the frequency dependence from the microscopic response kernel. Fig.~\ref{fig3}(d) then shows the temperature-dependent frequency response of $\overline{{J}}_{\mathrm{ph}}^{x}$ for the linearly polarized case, $\varphi=0$. At low temperature, the response retains two resonance-related features associated with the separated $\mathrm{A}_{1}(\mathrm{TO})$ and $\mathrm{E}_{1}(\mathrm{TO})$ modes. As temperature increases, anharmonic broadening smears these features and gradually merges them into a single broad response peak.
\par
The degenerate $\mathrm{E}_{1}(\mathrm{TO})$ channel exhibits a distinct helicity-selected response. As shown in Fig.~\ref{fig3}(e), $\overline{J}_{\mathrm{ph}}^{z}$ is concentrated around the degenerate $\mathrm{E}_{1}(\mathrm{TO})$ resonance and displays an approximately antisymmetric phase pattern with respect to $\varphi=0$, indicating that opposite field handedness generates angular momentum with opposite signs. This helicity selectivity is further confirmed in Fig.~\ref{fig3}(f): the linearly polarized drive is strongly suppressed, whereas the two circularly polarized drives produce responses with opposite signs. The peak magnitude is substantially larger than that of the nondegenerate $\mathrm{E}_{1}(\mathrm{TO})$--$\mathrm{A}_{1}(\mathrm{TO})$ channel, reflecting the resonant enhancement of the degenerate in-plane rotational motion. Fig.~\ref{fig3}(g) further highlights the strong frequency selectivity of this effect. Here $\omega_{+}$ and $\omega_{-}$ denote driving frequencies slightly detuned from the degenerate resonance, with detunings of $\pm 0.5$ THz in $\omega_{0}/2\pi$. Such a small detuning already substantially reduces the phase-dependent angular momentum, suggesting that the helicity-selected response appears as a pronounced resonance peak. Finally, Fig.~\ref{fig3}(h) shows that increasing temperature broadens the resonant response through anharmonic damping and reduces its peak magnitude, while preserving the helicity-selected phase structure of the degenerate in-plane channel.
\par
\begin{figure}[htbp]
\centering
\includegraphics[width=0.83\linewidth]{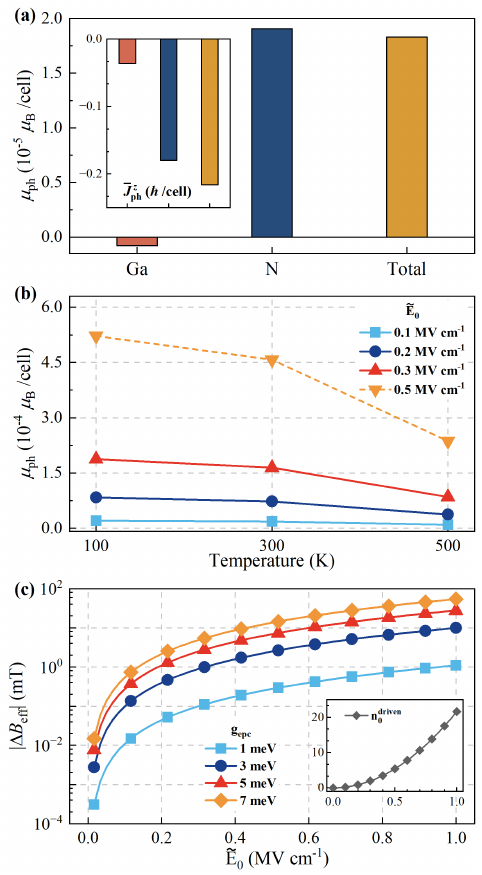}
\caption{Phonon orbital magnetic moment and effective magnetic-field estimate for the degenerate $\mathrm{E}_{1}(\mathrm{TO})$ channel. (a) Atom-resolved phonon magnetic moment $\mu_{\mathrm{ph}}$ induced by a Gaussian-enveloped THz pulse with $\tilde{E}_{0}=0.1~\mathrm{MV/cm}$ at 300 K; the inset shows the corresponding atom-resolved pulse-averaged angular momentum $\overline{J}^{z}_{\mathrm{ph}}$. (b) Temperature dependence of the total phonon orbital magnetic moment for different effective field amplitudes $\tilde{E}_{0}$. (c) Estimated effective magnetic field $\lvert \Delta B_{\mathrm{eff}}\lvert$ as a function of $\tilde{E}_{0}$ for representative electron--phonon coupling strengths $g_{\mathrm{epc}}$. The inset shows the corresponding effective number of circularly polarized phonons $n_{0}$.
}
\label{fig4}
\end{figure}
The sizable helicity-selected $\overline{J}_{\mathrm{ph}}^{z}$ suggests a possible magnetic response through dynamical multiferroicity, in which rotational ionic motion produces a magnetization along the rotation axis. In recent THz experiments on $\mathrm{SrTiO}_{3}$~\cite{basini2024terahertz}, magnetic responses associated with phonon angular momentum have been probed through time-resolved magneto-optical Kerr signals and interpreted microscopically within the phonon inverse Faraday-effect framework~\cite{PhysRevLett.133.266702}. In this framework, circularly polarized phonons couple to electronic degrees of freedom through electron--phonon interaction, giving rise to an electronic magnetization that can be expressed as an effective magnetic field. We therefore use $|\Delta B_{\mathrm{eff}}|$ to estimate the possible magnetic response associated with the induced phonon angular momentum and to assess its possible experimental observability. The corresponding evaluation procedure is described in App.~\ref{ap:C2}.
\par
The results at 300 K under resonant excitation with $\varphi=\pi/2$ are summarized in Fig.~\ref{fig4}. Fig.~\ref{fig4}(a) displays the atom-resolved phonon magnetic moment, with the corresponding atom-resolved $\overline{J}_{\mathrm{ph}}^{z}$ shown in the inset. Owing to the large mass contrast between Ga and N, together with their opposite effective charges, the ionic angular momentum is weighted very differently in the phonon magnetic moment. As a result, the N contribution dominates the total magnetic moment, whereas the Ga contribution is much smaller and has the opposite sign. Nevertheless, because the ionic masses are much larger than the electron mass, the bare phonon magnetic moment remains tied to the nuclear-magneton scale. The resulting phonon orbital magnetic moment is therefore only on the order of $10^{-5}\mu{\mathrm{B}}$ per unit cell, even though the induced phonon angular momentum reaches a sizable fraction of $\hbar$ per unit cell.
\par
Fig.~\ref{fig4}(b) further shows the temperature dependence of the total phonon orbital magnetic moment for several THz field amplitudes. Over the field range considered here, the magnetic moment increases from the $10^{-5}\mu_{\mathrm{B}}$ scale to the $10^{-4}\mu_{\mathrm{B}}$ scale per unit cell. The field dependence follows the expected quadratic trend of a second-order THz response, $\mu_{\mathrm{ph}}\propto \tilde{E}_{0}^{2}$, while the overall magnitude decreases with increasing temperature because anharmonic damping suppresses the resonant phonon response.
\par
Fig.~\ref{fig4}(c) presents the estimated effective magnetic field $|\Delta B_{\mathrm{eff}}|$ as a function of $\tilde{E}_{0}$ at 300 K. Following the phonon inverse Faraday-effect theory, we estimate this field from the electronic level splitting induced by circularly polarized phonons,
\begin{equation}
\lvert \Delta B_{\mathrm{eff}} \lvert
\simeq
\frac{\hbar\Omega_{0}\lvert g_{\mathrm{epc}}\lvert^2}{g_{J}\mu_{\mathrm{B}}(\epsilon^2_{g}-\hbar^2\Omega^2_{0})}n_{0}.
\end{equation}
Here $\Omega_{0}$ is the driven phonon angular frequency, $\epsilon_{g}$ is the electronic band gap, $g_{\mathrm{epc}}$ is the electron--phonon coupling strength, and $g_{J}$ is the electronic Landé factor. In the estimate, we take $\epsilon_{g}=3.40~\mathrm{eV}$ and $g_{J}=2$, while $g_{\mathrm{epc}}$ is varied from 1 to 7 meV to give a representative range of possible $|\Delta B_{\mathrm{eff}}|$ values for meV-scale coupling strengths. The $\mathrm{1/2}$ zero-point term in the original expression is omitted because we focus on the field-induced coherent phonon population. We therefore define $n_{0}=
\lvert\overline{J}^{z}_{\mathrm{ph}}\lvert/{\hbar}$ as the effective number of circularly polarized phonons associated with the induced angular momentum, rather than as an independently fitted parameter. With increasing $\tilde{E}_{0}$, $\lvert \Delta B_{\mathrm{eff}} \lvert$ increases rapidly and can reach the mT range for representative meV-scale electron--phonon coupling strengths. Although $\tilde{E}_{0}$ does not enter the effective-field expression explicitly, its influence is included through the field-induced circular-phonon number $n{0}$, as illustrated in the inset. This mT-scale estimate suggests that the magnetic response associated with the induced phonon angular momentum in GaN may lie within a potentially accessible experimental range.

\section{Conclusion}
In summary, this work establishes a microscopic response description of coherent THz-induced phonon angular momentum in polar crystals. The central result is a response-kernel formulation that connects the rotational lattice response to field-selected infrared-active modes, intermode angular-momentum matrix elements, and self-energy-dressed phonon propagators. In this form, the familiar driven-mode picture is retained, while finite-temperature frequency renormalization and linewidth broadening enter directly through the complex poles of the dressed Green's functions.
\par
The application to wurtzite GaN demonstrates how crystal anisotropy and phonon degeneracy shape THz-driven circular-phonon dynamics. Because the infrared-active TO modes and LO--TO splitting depend on the propagation direction, different transverse-field geometries select distinct rotational channels. The nondegenerate $\mathrm{E}_{1}(\mathrm{TO})$--$\mathrm{A}_{1}(\mathrm{TO})$ channel generates an $x$-directed angular momentum whose sign and magnitude can be tuned by the driving frequency and relative phase, including a finite response under linearly polarized excitation due to the intrinsic phase lag between the two modes. By contrast, the degenerate $\mathrm{E}_{1}(\mathrm{TO})$ channel is governed by an antisymmetric field combination and yields a helicity-selected $z$-directed response under circular or elliptical THz excitation. Anharmonic broadening suppresses and smooths the resonant structures in both channels, while preserving their characteristic phase-selection rules.
\par
The estimated magnetic-field scale further suggests that the THz-induced phonon angular momentum in GaN may lead to experimentally accessible magnetic signatures through a phonon inverse-Faraday-type mechanism. More broadly, the present formulation provides a controlled route beyond constant-damping oscillator models by incorporating mode selectivity, intermode angular-momentum matrix elements, and finite-temperature phonon spectral properties within a unified Green's-function framework. It also offers a starting point for future extensions that include fluctuation-level covariance contributions, vertex corrections, and angular-momentum transfer to electronic or spin degrees of freedom.

\begin{acknowledgments}
This work was supported by National Key R\&D Program of China (2023YFA1407001).
\end{acknowledgments}


\appendix
\section{Detailed derivation of the second-order response formalism}\label{ap:A}
In this Appendix, we provide the derivation of the second-order response formula used in Sec.~\ref{theory:A}. Starting from the interaction-picture expansion of the density matrix under the time-dependent electric-dipole perturbation, we derive the leading nonvanishing contribution to the expectation value of the phonon angular-momentum operator. This gives the causal time-domain response in Eq.~\eqref{eq:2ndresponse} and, after transforming to relative times and Fourier components, the dc response kernel in Eqs.~\eqref{eq:chikernel_dc} and \eqref{eq:Jph_dc}. The derivation follows the standard second-order Kubo construction and fixes the causal prescription used in the main text.
\subsection{Density-matrix expansion in the interaction picture}\label{ap:A1}
We start from a time-dependent total Hamiltonian
\begin{equation}
    \hat{\mathcal{H}}_{\mathrm{tot}}(t)=\hat{\mathcal{H}}+\hat{\mathcal{H}}^{\prime}(t)e^{st},
\end{equation}
where $\hat{\mathcal{H}}$ is the unperturbed lattice Hamiltonian and $\hat{\mathcal{H}}^{\prime}(t)=-\sum_{\mu}\tilde{E}_{\mu}(t)\hat{P}_{\mu}$ is the electric-dipole perturbation generated by the effective internal THz electric field $\tilde{\mathbf{E}}(t)$. The auxiliary factor $e^{st}$, with $s\to 0^{+}$, implements adiabatic switching from $t=-\infty$ and fixes the retarded causal prescription in the response functions. To derive the nonlinear response, we work in the interaction picture with respect to $\hat{\mathcal{H}}$, in which the density matrix is defined as
\begin{equation}
    \hat{\rho}_{I}(t)=e^{i\hat{\mathcal{H}}t/\hbar}\hat{\rho}(t)e^{-i\hat{\mathcal{H}}t/\hbar},
\end{equation}
and satisfies
\begin{equation}\label{eqa3}
    i\hbar\frac{\partial }{\partial t}\hat{\rho}_{I}(t)
    =e^{st}
    \[
    \hat{\mathcal{H}^{\prime}_{I}}(t),\hat{\rho}_{I}(t)
    \],
\end{equation}
where $\hat{\mathcal{H}^{\prime}_{I}}(t)=e^{i\hat{\mathcal{H}}t/\hbar}\hat{\mathcal{H}}^{\prime}(t)e^{-i\hat{\mathcal{H}}t/\hbar}$.
\par
We assume that at $t \to -\infty$ the system is described by the equilibrium density matrix $\hat{\rho}_{0}$ of the unperturbed lattice Hamiltonian, with $\[\hat{\mathcal{H}},\hat{\rho}_{0}\]=0$. The interaction-picture density matrix is then expanded in powers of the electric-field perturbation as~\cite{PhysRevB.110.245132}
\begin{equation}
    \hat{\rho}_{I}(t)=\hat{\rho}_{0}+\hat{\rho}_{I}^{(1)}(t)+\hat{\rho}_{I}^{(2)}(t)+\cdots,
\end{equation}
where the superscript denotes the perturbation order. Substituting this expansion into Eq.~\eqref{eqa3} and collecting equal powers of the perturbation gives
\begin{subequations}
\begin{align}
i\hbar\partial_{t}\hat{\rho}^{(1)}_{I}(t)&=
e^{st}[\mathcal{H}^{\prime}_{I}(t),\hat{\rho}_{0}],\\
i\hbar\partial_{t}\hat{\rho}^{(2)}_{I}(t)&=e^{st}[\mathcal{H}^{\prime}_{I}(t),\hat{\rho}^{(1)}_{I}(t)].
\end{align}
\end{subequations}
Formally integrating these equations from $t=-\infty$ to $t$, we obtain
\begin{equation}
    \hat{\rho}^{(1)}_{I}(t)=-\frac{i}{\hbar}\int_{-\infty}^{t}dt_{1}e^{st_{1}}
    [\mathcal{H}^{\prime}_{I}(t_1),\hat{\rho}_{0}],
\end{equation}
and
\begin{equation}\label{eqa7}
\begin{aligned}
    \hat{\rho}^{(2)}_{I}(t)&=
\(-\frac{i}{\hbar}\)^2
    \int_{-\infty}^{t}dt_{1}e^{st_{1}}
    \int_{-\infty}^{t_{1}}dt_{2}\\
    &\times e^{st_2}
\[
\hat{\mathcal{H}}^{\prime}_{I}(t_1),
[
\hat{\mathcal{H}}^{\prime}_{I}(t_2),\hat{\rho}_{0}
]
\].
\end{aligned}
\end{equation}
These expressions provide the perturbative density matrix used below to evaluate the leading nonvanishing contribution to the induced phonon angular momentum.

\subsection{Expectation value of phonon angular momentum}\label{ap:A2}
For a time-reversal-symmetric lattice in thermal equilibrium, phonon modes at $\mathbf{q}$ and $-\mathbf{q}$ carry opposite angular momenta and cancel pairwise in the Brillouin-zone summation~\cite{PhysRevLett.112.085503}. Therefore, the unperturbed lattice carries no net phonon angular momentum, and the zeroth-order contribution vanishes, 
\begin{equation}
\<\hat{J}^{\alpha}_{\mathrm{ph}}\>^{(0)}=0.
\end{equation}
Here and in the following, $\left<\cdots\right>\equiv \mathrm{Tr(\hat{\rho}_{0}\dots)}$ denotes the equilibrium average with respect to the unperturbed density matrix.
\par
Moreover, the phonon angular momentum operator is defined as~\cite{PhysRevLett.112.085503}
\begin{equation}
    \hat{\mathbf{J}}_{\mathrm{ph}}=\sum_{l\kappa}\hat{\mathbf{u}}_{l\kappa}\times m_{\kappa}\hat{\dot{{\mathbf{u}}}}_{l\kappa},
\end{equation}
which is bilinear in the lattice dynamical variables. By contrast, the electric field enters through the dipole coupling
\begin{equation}\label{eqa9}
    \hat{\mathcal{H}}^{\prime}(t)=-\sum_{\mu}\tilde{E}_{\mu}(t)\hat{P}_{\mu},
\end{equation}
where the polarization operator is linear in the ionic displacement. The corresponding linear-response term involves an equilibrium average of an operator that is linear in the lattice displacement or velocity and therefore vanishes in the expansion about the equilibrium structure. The first nonvanishing induced contribution to the phonon angular momentum is thus second order in the external electric field.
\par
Keeping the second-order term in the interaction-picture expansion of the density matrix derived in Sec.~\ref{ap:A1}, we obtain
\begin{equation}
{J}_{\mathrm{ph}}^{\alpha}(t)=
    \mathrm{Tr}\[
    \hat{\rho}_{I}(t)\hat{J}^{\alpha}_{I}(t)
    \]\simeq
    \mathrm{Tr}\[
    \hat{\rho}_{I}^{(2)}(t)\hat{J}^{\alpha}_{I}(t)
    \].
\end{equation}
Substituting Eq.~\eqref{eqa7} into this expression and using the cyclic invariance of the trace, we find
\begin{equation}
\begin{aligned}
{J}_{\mathrm{ph}}^{\alpha}(t)&=
\lim_{s\to0^{+}}
\(-\frac{i}{\hbar}\)^2
    \int_{-\infty}^{t}e^{st_{1}}dt_{1}\int_{-\infty}^{t_{1}}dt_2\\
    &\times e^{st_2}
    \left<\[
    \[
    \hat{J}_{I}^{\alpha}(t),\hat{\mathcal{H}}^{\prime}_{I}(t_1)
    \],
    \hat{\mathcal{H}}^{\prime}_{I}(t_2)
    \]\right>.
\end{aligned}
\end{equation}
Inserting the dipole perturbation \eqref{eqa9} then gives
\begin{equation}\label{eqa12}
\begin{aligned}
{J}_{\mathrm{ph}}^{\alpha}(t)&=\lim_{s\to 0^{+}}\sum_{\mu\nu}
\(-\frac{i}{\hbar}\)^2\int_{-\infty}^{t}e^{st_1}dt_{1}\int_{-\infty}^{t_{1}}e^{st_2}dt_{2}\\
&\times \tilde{E}_{\mu}(t_{1})
\tilde{E}_{\nu}(t_2)
\left<\[
    \[
    \hat{J}_{I}^{\alpha}(t),\hat{{P}}^{I}_{\mu}(t_1)
    \],
    \hat{{P}}^{I}_{\nu}(t_2)
    \]\right>.
\end{aligned}
\end{equation}
This expression makes explicit that the induced phonon angular momentum is bilinear in the electric field and is governed by a retarded double commutator involving the phonon-angular-momentum and polarization operators. In the next subsection, we transform this result to relative-time and frequency-domain forms to obtain the response kernel used in Sec.~\ref{theory:A}.

\subsection{Frequency-domain response kernel}\label{ap:A3}
To obtain the frequency-domain response kernel used in Sec.~\ref{theory:A}, we rewrite Eq.~\eqref{eqa12} in terms of relative times. We introduce
\begin{equation}
 u=t-t_1,\qquad
 v=t_1-t_2,
\end{equation}
so that $u\geq 0$, $v\geq 0$, and
\begin{equation}
    t_1=t-u,\qquad t_2=t-u-v.
\end{equation}
Using the time-translation invariance of the equilibrium average with respect to the unperturbed Hamiltonian, the double commutator $\< [
    [
    \hat{J}_{I}^{\alpha}(t),\hat{{P}}^{I}_{\mu}(t_1)
    ],
    \hat{{P}}^{I}_{\nu}(t_2)
    ]\>$ 
depends only on the time differences. Eq.~\eqref{eqa12} can therefore be written as
\begin{equation}
\begin{aligned}
{J}_{\mathrm{ph}}^{\alpha}(t)&=\lim_{s\to 0^{+}}\sum_{\mu\nu}
\int_{0}^{+\infty}du\int_{0}^{+\infty}dv \Pi_{\mu\nu}^{\alpha}(u,v)\\
&\times e^{s(2t-2u-v)}\tilde{E}_{\mu}(t-u)\tilde{E}_{\nu}(t-u-v),
\end{aligned}
\end{equation}
where
\begin{equation}\label{eqa17}
\Pi_{\mu\nu}^{\alpha}(u,v)=\(-\frac{i}{\hbar}\)^2 \left<[
    [
    \hat{J}_{I}^{\alpha}(u+v),\hat{{P}}^{I}_{\mu}(v)
    ],
    \hat{{P}}^{I}_{\nu}(0)
    ]\right>
\end{equation}
is the retarded time-domain response kernel.
\par
The adiabatic factor in the perturbation gives the convergence factor $e^{s(2t-2u-v)}$. Its $u$- and $v$-dependent part supplies the convergence factors for the causal relative-time integrals and leads, in the frequency domain, to the usual $+i0^+$ prescription. We use the Fourier-transform convention
\begin{equation}\label{eqA18}
\begin{aligned}
    \tilde{E}_{\mu}(t)&=\int_{-\infty}^{+\infty}\frac{d\omega}{2\pi}\tilde{E}_{\mu}(\omega)e^{-i\omega t},\\
    \mathcal{J}_{\mathrm{ph}}^{\alpha}(\Omega)&=\int_{-\infty}^{+\infty}dt e^{i\Omega t} {J}_{\mathrm{ph}}^{\alpha}(t).
\end{aligned}
\end{equation}
Substituting the Fourier components of the electric fields into the time-domain response and carrying out the integral over the observation time $t$, one obtains the bilinear frequency-domain form
\begin{equation} \label{eqA19}
\mathcal{J}^{\alpha}_{\mathrm{ph}}(\Omega)=\sum_{\mu\nu}
    \int_{-\infty}^{+\infty}d\omega
    \chi_{\mu\nu}^{\alpha}(\omega,\Omega-\omega)
    \tilde{E}_{\mu}(\omega)\tilde{E}_{\nu}(\Omega-\omega),
\end{equation}
where the response kernel is
    \begin{equation}\label{eqA20}
    \begin{aligned}
        \chi^{\alpha}_{\mu\nu}(\omega,\Omega-\omega)&=\frac{1}{2\pi}\int_{0}^{+\infty}du\int_{0}^{+\infty}dv
        \Pi^{\alpha}_{\mu,\nu}(u,v)\\
        &\times e^{i(\Omega+i0^{+})u}e^{i(\Omega-\omega+i0^{+})v}.
    \end{aligned}
    \end{equation}
In the main text, we set $\Omega=0$ to extract the zero-output-frequency component of the nonlinear response. Under the Fourier convention in Eq.~\eqref{eqA18}, this component corresponds to the zero-frequency Fourier weight of the rectified angular-momentum response. For the finite THz pulse considered in the numerical calculations, the corresponding pulse-averaged angular momentum is obtained by normalizing this zero-frequency weight by an effective pulse duration, as described in Sec.~\ref{Results} and App.~\ref{ap:C1}.

\section{Green's-function evaluation of the phonon-basis response kernel}\label{ap:B}

In this Appendix, we provide the Green's-function evaluation of the phonon-basis response kernel used in Secs.~\ref{theory:B} and \ref{theory:C}. Starting from the dc response kernel obtained in Eq.~\eqref{eq:chikernel_dc}, we express the phonon angular-momentum and polarization operators in terms of phonon normal modes. The resulting four-operator correlation functions are then reduced, within a bubble-level approximation, to products of retarded phonon Green's functions. We subsequently introduce the diagonal quasiparticle form of the self-energy-dressed propagator and evaluate the resulting frequency integrals by contour integration, leading to the compact linewidth-broadened kernel used in the main text. Finally, we summarize the reality property of the kernel and its narrow-linewidth limit, which connects the Green's-function expression to the familiar Lorentz-type resonant picture.
\subsection{Phonon-basis reduction of the kernel}\label{ap:B1}

We use the phonon-field operators introduced in Sec.~\ref{theory:B}, $\hat{\mathcal{A}}_{\mathbf{q}c}=\hat{a}_{\mathbf{q}c}+\hat{a}^{\dagger}_{-\mathbf{q}c}$ and $\hat{\mathcal{B}}_{\mathbf{q}c}=\hat{a}_{\mathbf{q}c}-\hat{a}^{\dagger}_{-\mathbf{q}c}$, where $\hat{a}_{\mathbf{q}c}$ and $\hat{a}^{\dagger}_{\mathbf{q}c}$ are the annihilation and creation operators of the phonon mode $(\mathbf{q},c)$, respectively. 
\par
In this representation, the phonon angular-momentum operator is written as
\begin{equation}\label{eqB1}
    \hat{J}_{I}^{\alpha}(t)=\sum_{\mathbf{q},ab}\mathcal{L}^{\alpha}_{\mathbf{q},ab}
    \hat{\mathcal{A}}^{I}_{\mathbf{q}a}(t)\hat{\mathcal{B}}^{I}_{-\mathbf{q}b}(t),
\end{equation}
with the mode-pair angular-momentum matrix element $\mathcal{L}^{\alpha}_{\mathbf{q},ab}$ defined by
\begin{equation}
    \mathcal{L}^{\alpha}_{\mathbf{q},ab}=-\frac{i\hbar}{2}\sqrt{\frac{\omega_{\mathbf{q}b}}{\omega_{\mathbf{q}a}}}
    \sum_{\kappa}
    \(
    \mathbf{e}_{\mathbf{q}a,\kappa}
    \times
    \mathbf{e}^{*}_{\mathbf{q}b,\kappa}
    \)_{\alpha}.
\end{equation}
Here, $\mathbf{e}_{\mathbf{q}a,\kappa}$ denotes the $\kappa$th atomic component of the phonon eigenvector for mode $(\mathbf{q},a)$ with frequency $\omega_{\mathbf{q}a}$. 
\par
Likewise, within the dipole approximation, the lattice-polarization operator is expanded as
\begin{equation}\label{eqB3}
    \hat{P}^{I}_{\mu}(t)=
    \sum_{a}\mathcal{P}^{\mu}_{a} \hat{\mathcal{A}}^{I}_{a}(t),
\end{equation}
with the mode-resolved polarization matrix element $\mathcal{P}^{\mu}_{a}$ given by
\begin{equation}\label{eqB4}
    \mathcal{P}^{\mu}_{a}=\sum_{\mathbf{q}}\sum_{\kappa\nu}
    \sqrt{\frac{\hbar N}{2m_{\kappa}\omega_{\mathbf{q}a}}}
    \mathcal{Z}^{*}_{\kappa,\mu\nu}e_{\mathbf{q}a,\kappa}^{\nu}\delta_{\mathbf{q,0}}.
\end{equation}
The factor $\delta_{\mathbf{q,0}}$ expresses the dipole selection rule for a spatially uniform THz electric field: only zone-center infrared-active modes couple directly to the field-induced lattice polarization.
\par
Substituting Eqs.~\eqref{eqB1} and \eqref{eqB3} into the time-domain kernel in Eq.~\eqref{eqa17} gives
\begin{widetext}
\begin{equation}\label{eqB5}
\begin{aligned}
    &\Pi_{\mu\nu}^{\alpha}(u,v)\\
    &=\(\frac{-i}{\hbar}\)^2
    \sum_{\mathbf{q}}\sum_{abcd}
    \mathcal{L}^{\alpha}_{\mathbf{q},ab}\mathcal{P}^{\mu}_{c}\mathcal{P}^{\nu}_{d}
    \left[
\left<\hat{\mathcal{A}}^{I}_{\mathbf{q}a}(u+v)
    \hat{\mathcal{B}}^{I}_{-\mathbf{q}b}(u+v)
    \hat{\mathcal{A}}^{I}_{c}(v)
    \hat{\mathcal{A}}^{I}_{d}(0)
    \right>
    -
    \left<
    \hat{\mathcal{A}}^{I}_{d}(0)
    \hat{\mathcal{A}}^{I}_{\mathbf{q}a}(u+v)
    \hat{\mathcal{B}}^{I}_{-\mathbf{q}b}(u+v)
    \hat{\mathcal{A}}^{I}_{c}(v)
    \right>
    \right]\\
    &+\(\frac{-i}{\hbar}\)^2\sum_{\mathbf{q}}\sum_{abcd}\mathcal{L}^{\alpha}_{\mathbf{q},ab}\mathcal{P}^{\mu}_{c}\mathcal{P}^{\nu}_{d}
    \left[
    \left< 
    \hat{\mathcal{A}}^{I}_{d}(0)
    \hat{\mathcal{A}}^{I}_{c}(v)
    \hat{\mathcal{A}}^{I}_{\mathbf{q}a}(u+v)
    \hat{\mathcal{B}}^{I}_{-\mathbf{q}b}(u+v)
    \right>
    -
    \left< 
    \hat{\mathcal{A}}^{I}_{c}(v)
    \hat{\mathcal{A}}^{I}_{\mathbf{q}a}(u+v)
    \hat{\mathcal{B}}^{I}_{-\mathbf{q}b}(u+v)
    \hat{\mathcal{A}}^{I}_{d}(0)
    \right>
    \right],
\end{aligned}
\end{equation}
\end{widetext}
which is the direct phonon-basis expansion of the retarded double commutator entering the second-order response kernel.
\par
To proceed, we apply a bubble-level decoupling to the four-operator correlation functions in Eq.~\eqref{eqB5}. For a generic product $\left<\hat{X}_1(t)\hat{X}_2(t)\hat{X}_3(t^{\prime})\hat{X}_4(t^{\prime\prime})\right>$, we retain the cross-vertex pairings
$
\left<\hat{X}_1\hat{X}_3\right>\left<\hat{X}_2\hat{X}_4\right>
+
\left<\hat{X}_1\hat{X}_4\right>\left<\hat{X}_2\hat{X}_3\right>
$, which connect the angular-momentum vertex to the two polarization insertions. The internal pairing $\left<\hat{X}_1\hat{X}_2\right>\left<\hat{X}_3\hat{X}_4\right>$ is not retained because it does not represent two field-induced propagation lines connecting the polarization vertices to the angular-momentum vertex. Connected four-point corrections are also omitted at this level.
\par
This approximation selects the coherent bubble contribution to the nonlinear response. It describes the part of the induced phonon angular momentum constructed from two field-driven single-phonon propagation processes. The omitted connected terms correspond to fluctuation-level or vertex-corrected contributions, such as field-induced changes in covariance-type phonon correlators, and are treated here as beyond the present self-energy-dressed bubble approximation.
\par
Applying this factorization to Eq.~\eqref{eqB5} and grouping the resulting terms according to their relative time arguments, the time-domain kernel is reduced to pairwise commutators of the forms $\left<\left[\hat{\mathcal{A}},\hat{\mathcal{A}}\right]\right>$ and $\left<\left[\hat{\mathcal{B}},\hat{\mathcal{A}}\right]\right>$, weighted by the angular-momentum and polarization matrix elements:
\begin{widetext}
\begin{equation}\label{eqB6}
\begin{aligned}
\Pi_{\mu\nu}^{\alpha}(u,v)
&=\(\frac{-i}{\hbar}\)^2
\sum_{\mathbf{q}}\sum_{abcd}\mathcal{L}^{\alpha}_{\mathbf{q},ab}\mathcal{P}^{\mu}_{c}\mathcal{P}^{\nu}_{d}
\left<
\left[
\hat{\mathcal{A}}^{I}_{\mathbf{q}a}(u),
\hat{\mathcal{A}}^{I}_{c}(0)
\right]
\right>
\left<
\left[
\hat{\mathcal{B}}^{I}_{-\mathbf{q}b}(u+v),
\hat{\mathcal{A}}^{I}_{d}(0)
\right]
\right>\\
&+\(\frac{-i}{\hbar}\)^2
\sum_{\mathbf{q}}\sum_{abcd}\mathcal{L}^{\alpha}_{\mathbf{q},ab}\mathcal{P}^{\mu}_{c}\mathcal{P}^{\nu}_{d}
\left<
\left[
\hat{\mathcal{A}}^{I}_{\mathbf{q}a}(u+v),
\hat{\mathcal{A}}^{I}_{d}(0)
\right]
\right>
\left<
\left[
\hat{\mathcal{B}}^{I}_{-\mathbf{q}b}(u),
\hat{\mathcal{A}}^{I}_{c}(0)
\right]
\right>.
\end{aligned}
\end{equation}
\end{widetext}
Since the response is causal and the relative times satisfy $u\ge 0$ and $v\ge 0$, these two-point commutators can be identified with retarded phonon Green's functions. We define
\begin{equation}
    G^{R}_{\lambda\lambda^{\prime}}(t)=\(-\frac{i}{\hbar}\)\theta(t)
    \left<
    \left[
    \hat{\mathcal{A}}_{\lambda}^{I}(t),
    \hat{\mathcal{A}}_{\lambda^{\prime}}^{I,\dagger}(0)
    \right]
    \right>,
\end{equation}
where $\lambda\equiv(\mathbf{q},a)$ is a composite mode index. Using $i\partial_{t}\hat{\mathcal{A}}^{I}_{\lambda}(t)=\omega_{\lambda}\hat{\mathcal{B}}^{I}_{\lambda}(t)$, the correlator involving the velocity-like operator can be rewritten as
\begin{equation}
    -\(\frac{i}{\hbar}\)\theta(t)
    \left<
    \left[
    \hat{\mathcal{B}}_{\lambda}^{I}(t),
    \hat{\mathcal{A}}_{\lambda^{\prime}}^{I,\dagger}(0)
    \right]
    \right>=
    \frac{i}{\omega_{\lambda}}
    \partial_{t}G^{R}_{\lambda\lambda^{\prime}}(t).
\end{equation}
The contact term from differentiating the step function vanishes because the equal-time commutator of the displacement-like field $\hat{\mathcal{A}}$ is zero. 
\par
With these relations, the decoupled kernel can be written entirely in terms of retarded phonon Green's functions as
\begin{equation}\label{eqB9}
\begin{aligned}
\Pi_{\mu\nu}^{\alpha}(u,v)
&= i \sum_{abcd}
\tilde{\mathcal{L}}^{\alpha}_{ab}\,
\mathcal{P}_{c}^{\mu}\mathcal{P}_{d}^{\nu}
\Big[
G^{R}_{ac}(u)\,
\partial_{u+v} G^{R}_{bd}(u+v)
\\
&\qquad\qquad
+\, G^{R}_{ad}(u+v)\,
\partial_{u} G^{R}_{bc}(u)
\Big].
\end{aligned}
\end{equation}
Here, translational symmetry makes the retarded phonon Green’s function diagonal in crystal momentum, while time-reversal symmetry gives $\omega_{-\mathbf{q}a}=\omega_{\mathbf{q}a}$. Together with the dipole selection rule in Eq.~\eqref{eqB4}, this restricts the field-coupled response to the zone-center sector. We therefore omit the wave-vector index in Eq.~\eqref{eqB9} and introduce 
\begin{equation}
\tilde{\mathcal{L}}_{ab}^{\alpha}\equiv\mathcal{L}_{ab}^{\alpha}/\omega_{b},
\end{equation}
which satisfies $\tilde{\mathcal{L}}^{\alpha}_{ab}=\tilde{\mathcal{L}}^{\alpha,*}_{ba}$. 
\par
By substituting the Fourier representation of the retarded phonon Green’s function,
\begin{equation}
    G^{R}_{ab}(t)=\int_{-\infty}^{+\infty}\frac{d\omega}{2\pi}G^{R}_{ab}(\omega)e^{-i\omega t}
\end{equation}
into Eq.~\eqref{eqB9}, and using the dc limit of Eq.~\eqref{eqA20}, we obtain
\begin{equation}\label{eqB11}
\begin{aligned}
&\chi_{\mu\nu}^{\alpha}(\omega,-\omega)\\
&=
-\sum_{abcd}
\frac{\tilde{\mathcal{L}}^{\alpha}_{ab}\mathcal{P}^{\mu}_{c}\mathcal{P}^{\nu}_{d}}{(2\pi)^3}
\int_{-\infty}^{+\infty} d\omega_{1}
\int_{-\infty}^{+\infty} d\omega_{2}
\\
&\quad\times
\frac{
\omega_{2}G^{R}_{ac}(\omega_1)G^{R}_{bd}(\omega_2)
+\omega_{1}G^{R}_{bc}(\omega_1)G^{R}_{ad}(\omega_2)
}{
(\omega_1+\omega_2-i0^{+})
(\omega+\omega_2-i0^{+})
},
\end{aligned}
\end{equation}
which corresponds to Eq.~\eqref{eq:chikernel_gr} in the main text. The two denominators in Eq.~\eqref{eqB11} originate from the integrations over the positive relative times $u$ and $v$, and encode the causal structure of the second-order response. This expression organizes the dc kernel as a convolution of two retarded phonon propagators. The frequency dependence of the propagators determines the resonant structure, while the polarization and angular-momentum matrix elements set the weight of each mode-pair channel. This Green's-function representation thus provides the starting point for introducing anharmonic self-energy corrections and deriving the linewidth-broadened response kernel in the next subsection.

\subsection{Residue-theorem derivation of the reduced response kernel}\label{ap:B2}
Since the self-energy dressing of retarded phonon Green's functions is standard in anharmonic lattice dynamics, we do not repeat its derivation here. Instead, we use the dressed Green's function introduced in Sec.~\ref{theory:C} and focus on the contour evaluation of the Green's-function convolution in Eq.~\eqref{eqB11}. For the well-defined zone-center optical modes considered in this work, we retain only the diagonal part of the anharmonic self-energy in the phonon eigenmode basis. In this basis, the off-diagonal components $G^{R}_{ab}$ with $a\neq b$ have no zeroth-order contribution and arise only from off-diagonal self-energy insertions in the Dyson equation. They are therefore neglected within the present diagonal quasiparticle approximation, so that $G^{R}_{ab}(\omega)\simeq \delta_{ab}G^{R}_{a}(\omega)$. Under this approximation, Eq.~\eqref{eqB11} reduces to 
\begin{widetext}
\begin{equation}\label{eqB12}
\chi_{\mu\nu}^{\alpha}(\omega,-\omega)
=
-\sum_{ab}
\frac{{\tilde{\mathcal{L}}}^{\alpha}_{ab}}{(2\pi)^3}
\int_{-\infty}^{+\infty} d\omega_1
\int_{-\infty}^{+\infty} d\omega_2 
\frac{\Big[
\omega_2
\mathcal{P}^{\mu}_{a}\mathcal{P}^{\nu}_{b}
{G}^{R}_{a}(\omega_1){G}^{R}_{b}(\omega_2)
+
\omega_1
\mathcal{P}^{\mu}_{b}\mathcal{P}^{\nu}_{a}
{G}^{R}_{b}(\omega_1){G}^{R}_{a}(\omega_2)
\Big]}{
(\omega_1+\omega_2-i0^{+})(\omega+\omega_2-i0^{+})
}
.
\end{equation}
\end{widetext}
\par
In the quasiparticle regime, where the dressed phonon poles remain well defined and the self-energy varies smoothly near each pole, the retarded propagator can be represented by the pole form in Eq.~\eqref{eq13}. Substituting this form into Eq.~\eqref{eqB12}, the remaining frequency integrals can be evaluated analytically by the residue theorem. For compact notation, we denote the complex pole positions by 
\begin{equation}
\zeta^{\eta_{a}}_{a}=\eta_{a}\tilde{\omega}_{a}-i\Gamma_{a},
\qquad
\eta_{a}=\pm1,
\end{equation}
so that each dressed retarded Green's function is decomposed into its positive- and negative-frequency pole contributions. The response kernel is then expressed as a sum of rational functions whose singularities are determined by the complex poles of the Green's function.
\par
The contour integrations are carried out sequentially. We first perform the $\omega_1$ integration by closing the contour in the lower half-plane, consistent with the retarded pole structure. For the term proportional to $\omega_2G^{R}_{a}(\omega_1)G^{R}_{b}(\omega_2)$ in Eq.~\eqref{eqB12}, the enclosed phonon pole is $\omega_1 = \zeta_{a}^{\eta_{a}}$. The causal denominator $(\omega_1+\omega_2-i0^{+})^{-1}$ gives a pole at $\omega_1=-\omega_2+i0^{+}$, 
which lies in the upper half-plane and therefore does not contribute. The remaining $\omega_2$ integral contains the retarded phonon pole $\omega_2=\zeta_{b}^{\eta_b}$, whereas the poles at $\omega_{2}=-\zeta_{a}^{\eta_{a}}+i0^{+}$ and $\omega_2 = -\omega+i0^{+}$ lie in the upper half-plane and are excluded. This gives a contribution proportional to
\begin{equation}
    \frac{\zeta_{b}^{\eta_{b}}}
    {\zeta_{a}^{\eta_{a}}+\zeta_{b}^{\eta_{b}}}
    \frac{1}{\omega+\zeta_{b}^{\eta_{b}}}.
\end{equation}
\par
The term proportional to $\omega_1G^{R}_{b}(\omega_1)G^{R}_{a}(\omega_2)$ is evaluated analogously. In this case, the $\omega_1$ integration picks the pole $\omega_1 = \zeta_{b}^{\eta_{b}}$, and the subsequent $\omega_2$ integration picks the pole $\omega_2 = \zeta_{a}^{\eta_{a}}$. The polarization factors are interchanged, and the final resonant denominator becomes $\omega+\zeta_{a}^{\eta_a}$. This gives a contribution proportional to
\begin{equation}
    \frac{\zeta_{b}^{\eta_{b}}}
    {\zeta_{a}^{\eta_{a}}+\zeta_{b}^{\eta_{b}}}
    \frac{1}{\omega+\zeta_{a}^{\eta_{a}}}.
\end{equation}
In the two expressions above, only the pole-dependent factors are displayed; the common pole weights, including the factor $(\eta_{a}\eta_{b}/\hbar^2)$, have been suppressed for clarity.
\par
Combining the two contributions, restoring the pole residues and polarization prefactors, and summing over all pole configurations $(\eta_a,\eta_b)$, we obtain the compact linewidth-broadened response kernel in Eqs.~\eqref{eq14} and \eqref{eq15}. This derivation shows explicitly that the broadened response is governed by the complex pole structure of the self-energy-dressed retarded phonon Green's functions.

\subsection{Reality condition and narrow-linewidth limit of the reduced kernel}\label{ap:B3}
We next summarize two general consequences of the reduced kernel in Eqs.~\eqref{eq14} and \eqref{eq15}: the reality condition required for the induced angular momentum and the narrow-linewidth limit that connects the Green's-function expression to the Lorentz-type resonant picture.
\par
We first examine the conjugation property of the reduced kernel. For a time-reversal-symmetric lattice, the zone-center phonon eigenvectors can be chosen real in the phonon basis used here. In this basis, the polarization matrix elements are real, while the phonon angular-momentum matrix elements are purely imaginary,
\begin{equation}
 \tilde{\mathcal{L}}^{\alpha *}_{ab}=-\tilde{\mathcal{L}}^{\alpha}_{ab}.
\end{equation}
At the same time, the complex poles of the dressed retarded Green's function satisfy
\begin{equation}
    (\zeta_{a}^{\eta_{a}})^{*}=-\zeta_{a}^{-\eta_{a}}.
\end{equation}
Using this relation in Eq.~\eqref{eq15}, and relabeling the summation indices as $\eta_{a}\to-\eta_{a}$ and $\eta_{b}\to-\eta_{b}$, one obtains
\begin{equation}
    \mathcal{K}_{ab}^{\mu\nu}(-\omega)=
    -\[
     \mathcal{K}_{ab}^{\mu\nu}(\omega)
    \]^{*}.
\end{equation}
Combining this result with Eq.~\eqref{eq14} and the purely imaginary character of $\tilde{\mathcal{L}}^{\alpha}_{ab}$, we obtain
\begin{equation}
    \chi_{\mu\nu}^{\alpha}(-\omega,\omega)=[\chi^{\alpha}_{\mu\nu}(\omega,-\omega)]^{*}.
\end{equation}
This is the expected reality condition of the second-order response kernel. It guarantees that, for a physical driving field satisfying $\tilde{E}_{\mu}(-\omega)=\tilde{E}^{*}_{\mu}(\omega)$, the induced phonon angular momentum obtained from Eq.~\eqref{eq:Jph_dc} is real.
\par
We now turn to the narrow-linewidth limit. When the dressed phonon poles remain well separated and the damping rates satisfy
\begin{equation}
    \Gamma_{a},\Gamma_{b} \ll \tilde{\omega}_{a},\tilde{\omega}_{b},
\end{equation}
the reduced kernel can be simplified by replacing the complex pole positions in the numerators and nonresonant mixed denominators by their real parts, while retaining the infinitesimal imaginary parts in the resonant denominators. From Eq.~\eqref{eq15}, this gives
\begin{equation}
\begin{aligned}
    \mathcal{K}_{ab}^{\mu\nu}(\omega)
    &\simeq
    \sum_{\eta_{a},\eta_{b}=\pm 1}
    \frac{
    \eta_{a}\tilde{\omega}_{b}
    }{\eta_{a}\tilde{\omega}_{a}+
    \eta_{b}\tilde{\omega}_{b}}
    \\
    &\times\(
    \frac{\mathcal{P}_{a}^{\mu}\mathcal{P}_{b}^{\nu}}{\omega+\eta_{b}\tilde{\omega}_{b}-i0^{+}}
    +
    \frac{\mathcal{P}_{b}^{\mu}\mathcal{P}_{a}^{\nu}}{\omega+\eta_{a}\tilde{\omega}_{a}-i0^{+}}
    \).
\end{aligned}
\end{equation}
In this limit, the response is resolved into four discrete pole channels associated with the positive- and negative-frequency poles of the two dressed propagators. The kernel therefore reduces to a sum of sharply defined resonant contributions centered at $\omega=\pm\tilde{\omega}_{a}$ and $\omega=\pm\tilde{\omega}_{b}$. Using the Sokhotski--Plemelj formula, each resonant denominator can be decomposed into a principal-value part and a $\delta$-function contribution, showing explicitly that the spectral weight collapses onto discrete resonances as the linewidth tends to zero.
\par
This limit makes the connection to the conventional Lorentz-type picture transparent. When the linewidths are sufficiently small, the Green's-function kernel reduces to a discrete set of resonant mode-pair channels. Away from this limit, the full expression in Eqs.~\eqref{eq14} and \eqref{eq15} shows that the response is governed by the complex poles of the anharmonically dressed propagators, so that frequency renormalization and finite linewidths reshape the relative weights of different mode-pair contributions beyond a prescribed constant-damping description.

\section{Pulse-response evaluation and magnetic-field estimate}\label{ap:C}
In this Appendix, we provide the numerical details underlying the pulse-response calculations presented in Sec.~\ref{Results}. We first specify the Gaussian-enveloped THz pulse and its Fourier components under the convention of Eq.~\eqref{eqA18}. We then show how the field products are contracted with the response kernel for the nondegenerate and degenerate phonon channels. Finally, we summarize the evaluation of the phonon orbital magnetic moment and the effective magnetic field associated with the helicity-selected circular-phonon response.

\subsection{Numerical evaluation under Gaussian THz pulses}\label{ap:C1}
In the main text, the driving field is introduced as the effective internal THz electric field acting on the lattice polarization. Here we specify the considerations used to choose its representative amplitude. This choice is intended as an order-of-magnitude estimate rather than a full electrodynamic modeling of THz propagation inside GaN. Terahertz dielectric measurements of wurtzite GaN give static dielectric constants $\epsilon_{0,\perp}\approx 9.22$ and $\epsilon_{0,\parallel}\approx 10.32$~\cite{HIBBERD201668}. These values indicate a substantial dielectric screening of the internal field relative to the incident field. Following the dielectric-screening estimate commonly used in dynamical-multiferroicity calculations, we take $\tilde{E}_{0}= 0.1$ MV/cm as a representative effective internal-field amplitude. This value corresponds to an incident-field scale on the order of MV/cm, although the precise conversion depends on the frequency-dependent dielectric response, absorption near the infrared-active phonon resonance, and sample geometry. This field scale is comparable to those used in recent coherent-phonon THz experiments~\cite{kozina2019terahertz,juraschek2025chiral}. Based on these considerations, we use $\tilde{E}_{0}=0.1$ MV/cm in the main numerical calculations unless otherwise specified.
\par
With this convention, the driving field is modeled as a Gaussian-enveloped THz pulse with two orthogonal transverse components. For a generic pair of field directions $\mu$ and $\nu$, the time-domain field is written as
\begin{subequations}
\begin{align}
\tilde{E}_{\mu}(t)&=\tilde{E}_{0}g(\tau)\cos(\omega_{0}\tau),\\
\tilde{E}_{\nu}(t)&=\tilde{E}_{0}g(\tau)\cos(\omega_{0}\tau+\varphi),
\end{align}
\end{subequations} 
where $\tau=t-t_{0}$ is the time measured from the pulse center, $\omega_{0}$ is the central driving angular frequency, and $\varphi$ is the phase difference between the two components. The envelope function is taken as
\begin{equation}
g(\tau) = \exp \left(-\frac{\tau^2}{2\sigma^2}\right),
\end{equation}
with $\sigma = {\tau_{\mathrm{FWHM}}}/(2\sqrt{2\ln 2})$. In the numerical calculations shown in the main text, the pulse-envelope FWHM is set to $\tau_{\mathrm{FWHM}}=1.5$ ps.
\par
Using the Fourier-transform convention in Eq.~\eqref{eqA18}, the corresponding frequency-domain fields are
\begin{subequations}
\begin{align}
\tilde{E}_{\mu}(\omega)&=\mathcal{E}_{0}e^{i\omega t_{0}}\left[F_{+}(\omega)+F_{-}(\omega)\right],\\
\tilde{E}_{\nu}(\omega)&=\mathcal{E}_{0}e^{i\omega t_{0}}\left[e^{-i\varphi}F_{+}(\omega)+e^{i\varphi}F_{-}(\omega)\right],
\end{align}
\end{subequations}
where
\begin{equation}
\mathcal{E}_{0}=\frac{\tilde{E}_{0}\sigma \sqrt{2\pi}}{2},\qquad
F_{\pm}(\omega) = \exp\[
-\frac{\sigma^2}{2}(\omega\mp\omega_{0})^2
\].
\end{equation}
The field product entering the zero-output-frequency response is then
$
A_{\mu\nu}(\omega)\equiv \tilde{E}_{\mu}(\omega)\tilde{E}_{\nu}(-\omega)
$. Since the time-domain electric field is real, $\tilde{E}_{\mu}(-\omega)=\tilde{E}^{*}_{\mu}(\omega)$. It follows that $A_{\mu\nu}(-\omega)=A^{*}_{\mu\nu}(\omega)$, with
\begin{subequations}
\begin{align}
\mathrm{Re}A_{\mu\nu}(\omega)&=\mathcal{E}^2_{0}\cos\varphi \[F_{+}(\omega)+F_{-}(\omega)\]^2,\\
\mathrm{Im}A_{\mu\nu}(\omega)&=\mathcal{E}^2_{0}\sin\varphi \[F^2_{+}(\omega)-F^2_{-}(\omega)\].\label{eqC5b}
\end{align}
\end{subequations}
Thus, $\mathrm{Re}A_{\mu\nu}(\omega)$ is even in frequency, whereas $\mathrm{Im}A_{\mu\nu}(\omega)$ is odd. The phase difference $\varphi$ therefore controls the relative weights of the symmetric and antisymmetric field contractions in the rectified response.
\par
For the nondegenerate $\mathrm{E}_{1}(\mathrm{TO})$--$\mathrm{A}_{1}(\mathrm{TO})$ channel, the two transverse field components are taken along $y$ and $z$, and the induced angular momentum is along the $x$ direction. The relevant response-kernel components are $\chi_{yz}^{x}$ and $\chi_{zy}^{x}$, which are not constrained to be antisymmetric. The zero-output-frequency angular-momentum component generated by the pulse is therefore
\begin{equation}\label{eqC6}
\begin{aligned}
\mathcal{J}_{\mathrm{ph,0}}^{x}&=
\int_{-\infty}^{+\infty}d\omega \tilde{E}_{y}(\omega)\tilde{E}_{z}(-\omega)\chi_{yz}^{x}(\omega)\\
&+\int_{-\infty}^{+\infty}d\omega \tilde{E}_{z}(\omega)\tilde{E}_{y}(-\omega)\chi_{zy}^{x}(\omega)\\
&=\int_{-\infty}^{+\infty}d\omega A_{yz}(\omega)B_{yz}^{x}(\omega),
\end{aligned}
\end{equation}
where for brevity, $\chi_{\mu\nu}^{x}(\omega)\equiv \chi_{\mu\nu}^{x}(\omega,-\omega)$. In the last equality, the second term has been rewritten by changing $\omega\to-\omega$ and using the reality condition of the response kernel. The auxiliary function $B^{x}_{yz}(\omega)$ satisfies
\begin{subequations}
\begin{align}
\mathrm{Re} B_{yz}^{x}(\omega)=\mathrm{Re}\chi_{yz}^{x}(\omega)+\mathrm{Re}\chi^{x}_{zy}(\omega),\\
\mathrm{Im} B_{yz}^{x}(\omega)=\mathrm{Im}\chi_{yz}^{x}(\omega)-\mathrm{Im}\chi^{x}_{zy}(\omega).
\end{align}
\end{subequations}
The same reality condition implies that $\mathrm{Re} B_{yz}^{x}(\omega)$ is even in frequency, whereas $\mathrm{Im} B_{yz}^{x}(\omega)$ is odd. Together with the parity of $A_{yz}(\omega)$, the response can be evaluated over positive frequencies as
\begin{equation}
\begin{aligned}
\mathcal{J}_{\mathrm{ph,0}}^{x}&
=2\int_{0}^{+\infty}d\omega \mathrm{Re}A_{yz}(\omega)\mathrm{Re} B_{yz}^{x}(\omega)\\
&-2\int_{0}^{+\infty}d\omega \mathrm{Im}A_{yz}(\omega)\mathrm{Im} B_{yz}^{x}(\omega).
\end{aligned}
\end{equation}
This expression shows that the nondegenerate channel contains both a $\cos\varphi$-dependent contribution and a $\sin\varphi$-dependent contribution. As a result, a finite $\mathcal{J}_{\mathrm{ph,0}}^{x}$ can be generated even for linearly polarized THz excitation, because the two nondegenerate driven modes acquire an intrinsic relative phase lag through their unequal response functions.
\par
We next consider the degenerate $\mathrm{E}_{1}(\mathrm{TO})$ channel. In this case, the THz field is confined to the basal plane, and the induced angular momentum is along the $z$ direction. The relevant response components satisfy the antisymmetric relation $\chi_{xy}^{z}(\omega)=-\chi_{yx}^{z}(\omega)$, together with the reality condition $\chi_{xy}^{z}(-\omega)=[\chi_{xy}^{z}(\omega)]^{*}$. The zero-output-frequency angular-momentum component is then
\begin{equation}\label{eqC9}
\begin{aligned}
\mathcal{J}_{\mathrm{ph,0}}^{z}
&=\int_{-\infty}^{+\infty}d\omega \tilde{E}_{x}(\omega)\tilde{E}_{y}(-\omega)\chi_{xy}^{z}(\omega)\\
&+\int_{-\infty}^{+\infty}d\omega \tilde{E}_{y}(\omega)\tilde{E}_{x}(-\omega)\chi_{yx}^{z}(\omega)\\
&=-4\int_{0}^{+\infty}d\omega \mathrm{Im}A_{xy}(\omega)\mathrm{Im}\chi_{xy}^{z}(\omega).
\end{aligned}
\end{equation}
Using Eq.~\eqref{eqC5b}, the last line makes the helicity selection of the degenerate in-plane channel explicit. For linearly polarized driving, $\varphi=0$ or $\pi$, the antisymmetric field factor $\mathrm{Im}A_{xy}(\omega)$ vanishes, and this channel does not contribute to $\mathcal{J}^{z}_{\mathrm{ph,0}}$. For circularly polarized driving, $\varphi = \pm \pi/2$, the response is maximized and reverses sign when the helicity of the THz field is reversed.
\par
Equations above give the zero-output-frequency Fourier weight generated by the Gaussian THz pulse. Under the Fourier convention in Eq.~\eqref{eqA18}, this quantity is related to the time-domain angular momentum by
\begin{equation}
\mathcal{J}_{\mathrm{ph},0}^{\alpha}\equiv 
\mathcal{J}^{\alpha}_{\mathrm{ph}}(\Omega=0)
=
\int_{-\infty}^{+\infty}dt
{J}_{\mathrm{ph}}^{\alpha}(t).
\end{equation}
Thus, $\mathcal{J}^{\alpha}_{\mathrm{ph},0}$ represents the time-integrated weight of the rectified angular-momentum response. Because the Gaussian THz pulse is temporally localized, the induced rectified response is also confined to a finite effective time interval around the pulse. To report this pulse-integrated quantity as a pulse-averaged angular momentum, we normalize it by an effective averaging time,
\begin{equation}
\overline{J}^{\alpha}_{\mathrm{ph}}= \frac{1}{\tau_{\mathrm{av}}}\int_{-\infty}^{+\infty}dt
{J}_{\mathrm{ph}}^{\alpha}(t)=\frac{\mathcal{J}_{\mathrm{ph},0}^{\alpha}}{\tau_{\mathrm{av}}}.
\end{equation}
In the numerical evaluation, we take $\tau_{\mathrm{av}}=2\tau_{\mathrm{FWHM}}$. For the pulse parameters used in the main-text calculations, $\tau_{\mathrm{FWHM}}=1.5$ ps, giving $\tau_{\mathrm{av}}=3.0$ ps. This choice fixes the normalization of the reported pulse-averaged angular momentum; changing $\tau_{\mathrm{av}}$ would rescale the magnitude but would not affect the frequency, phase, or temperature dependences discussed in the main text.

\subsection{Phonon magnetic moment and effective-field estimate}\label{ap:C2}
We finally specify the definition of the phonon magnetic moment adopted in this work and the parameter choices entering the effective-field estimate. Since the largest induced angular momentum is obtained for the degenerate $\mathrm{E}_{1}(\mathrm{TO})$ pair, we focus on the $z$-directed angular-momentum response generated by resonant circularly polarized THz excitation.
\par
In this work, the phonon magnetic moment is evaluated from the ionic gyromagnetic relation between angular momentum and magnetic moment. For the in-plane circular motion, the total orbital phonon magnetic moment per unit cell is defined as~\cite{PhysRevMaterials.3.064405}
\begin{equation}
\mu_{\mathrm{ph}}^{z} = \sum_{\kappa}
\frac{e\mathcal{Z}^{*}_{\kappa,\perp}}{2m_{\kappa}}\overline{J}^{z}_{\kappa,\mathrm{ph}},
\end{equation}
where $e$ is the positive elementary charge, $\kappa$ labels the atom in the unit cell, $\mathcal{Z}^{*}_{\kappa,\perp} =\mathcal{Z}^{*}_{\kappa,xx}=\mathcal{Z}^{*}_{\kappa,yy}$ is the in-plane Born effective charge, and $\overline{J}^{z}_{\kappa,\mathrm{ph}}$ is the atom-resolved pulse-averaged angular momentum.
\par
By decomposing the angular-momentum matrix element into atomic contributions, $\mathcal{L}_{ab}^{z}=\sum_{\kappa}\mathcal{L}_{\kappa,ab}^{z}$, the atom-resolved angular momentum is obtained from the same response-kernel calculation, and the total pulse-averaged angular momentum then satisfies
\begin{equation}
\overline{J}_{\mathrm{ph}}^{z}=\sum_{\kappa}\overline{J}^{z}_{\kappa,\mathrm{ph}}.
\end{equation}
This definition keeps the sublattice-dependent charge-to-mass ratio explicitly in the magnetic moment. Thus, although the total angular momentum is obtained by summing the atom-resolved angular momenta, the total magnetic moment is a weighted sum in which different atoms carry different gyromagnetic factors. In GaN, the opposite Born effective charges of Ga and N, together with their large mass contrast, lead to distinct sublattice contributions to $\mu_{\mathrm{ph}}^{z}$.
\par
For the effective magnetic field, we use the phonon inverse-Faraday-type expression introduced in the main text. To estimate the possible magnetic-field scale, the driven phonon frequency is taken as the resonant frequency of the degenerate $\mathrm{E}_{1}(\mathrm{TO})$ mode, and the effective circular-phonon number is defined as $n_{0}=\lvert\overline{J}^{z}_{\mathrm{ph}} \lvert/\hbar$. The electronic band gap is set to 3.4 eV, consistent with the wide-gap character of wurtzite GaN, and the electronic Landé factor is set to $2$, providing an order-of-magnitude conversion between the circular-phonon-induced electronic level splitting and an effective magnetic field. The electron--phonon coupling strength is varied from $1$ to $7$ meV to give a representative range of $\lvert \Delta B_{\mathrm{eff}}\lvert$ for meV-scale coupling strengths.
\par
These magnetic-moment and effective-field estimates translate the calculated THz-induced phonon angular momentum into experimentally relevant magnetic scales, providing possible signatures of the predicted circular-phonon response.

\normalem
%


\end{document}